\documentclass[11pt]{article}
\usepackage{url}
\usepackage{comment}
\usepackage{algorithm}
\usepackage{algpseudocode}
\usepackage{xcolor}
\usepackage{amsthm}
\usepackage{amsmath}

\usepackage[margin=1in]{geometry}

\usepackage{amssymb, amsfonts, amsthm}

\usepackage{graphicx}
\usepackage{booktabs}
\usepackage{multirow}

\usepackage{natbib}

\usepackage{microtype}
\usepackage{setspace}

\newtheorem{theorem}{Theorem}
\newtheorem{proposition}[theorem]{Proposition}
\newtheorem{remark}[theorem]{Remark}

\begin{document}

% Title of paper
\title{Two-stage Adaptive Testing of Large-scale Mediation Hypotheses}

% List of authors, with corresponding author marked by asterisk

\author{Yueqi Xu \and \ Kwun Chuen Gary Chan \\[4pt]
% Author addresses
\textit{Department of Biostatistics, University of Washington,
Seattle, Washington 98195, U.S.A.}
\\[2pt]
% E-mail address for correspondence
{xyq678@uw.edu}}

% Running headers of paper:
\markboth%
% First field is the short list of authors
{Y. Xu and K.C.G. Chan}
% Second field is the short title of the paper
{Two-stage Adaptive Mediation Testing}

\maketitle

% Add a footnote for the corresponding author if one has been
% identified in the author list

\begin{abstract}
{In testing large-scale mediation hypotheses, the exposure-mediator and mediator-outcome path-specific p-values obtained for each hypothesis from the same sample can be combined into a pair of asymptotically independent p-values, which can then be used to test a global null hypothesis and a composite mediation null hypothesis, respectively. A first-stage screening procedure targeting the more stringent global null can effectively reduce the number of mediation hypotheses to be tested in the second stage, which in turn reduces the conservativeness of multiple comparison.  The framework can incorporate any data-adaptive choice of screening threshold in Stage 1, and any step-up false discovery rate control method in Stage 2. The proposed procedure controls the false discovery rate asymptotically while being consistently well powered across a wide range of scenarios. }
{False Discovery Rate; Mediation testing; Large-scale inference; Multiple testing}
\end{abstract}

\section{Introduction}
\label{sec: intro}

A treatment or exposure may have an effect on an outcome of interest through different pathways, including those operating through intermediate variables. In mediation analysis, a mediator is defined as a variable that is causally affected by the treatment and, in turn, causally affects the outcome (\citealp{baron1986moderator}). % Let $A$ denote the treatment/exposure, $M$ be the mediator, and $Y$ the outcome of interest, then $M$ falls in the causal pathway $A\longrightarrow M\longrightarrow Y$ (\citealp{baron1986moderator}). 
Unlike pre-treatment baseline confounders, which are measured prior to exposure and may influence the exposure, mediators are post-treatment, meaning their values can change depending on the treatment received, and mediation analysis often aims to quantify the extent of the total effect of an exposure on an outcome is being transmitted through such an induced change in a mediator.

As the advancement of epigenome-wide and other high-throughput molecular studies, there has been growing interest in assessing whether measured molecular features mediate the association between an exposure and a health outcome. For example, epigenome-wide mediation analyses have been used to screen DNA methylation CpG sites to identify loci that may mediate the effect of air pollution on coagulation and inflammation in the Normative Aging Study (\citealp{bind2014air}). Besides, large-scale mediation analysis has also been applied to identify gene expressions that mediate exposure effects (e.g., genetic variants) on downstream molecular outcomes (\citealp{yang2017identifying}). In such cases, instead of a single mediator, there are potentially thousands to hundreds of thousands of candidate mediators being tested simultaneously, and the key challenge is to correctly identify the subset that carry mediation effects while controlling the false discovery rate (FDR).

Let $A$ denote the treatment or exposure, $M_j$ the $j$-th candidate mediator for $j=1,\ldots, J$, $Y$ the outcome of interest, and $\mathbf X$ a vector of baseline confounders. For each potential mediator $M_j$, conventional mediation testing fits the following sets of linear models: 
\begin{equation}
\label{Eq: mediator_models}
\begin{aligned}
M_j &= \delta_{0,j} + \delta_j A + \boldsymbol\delta_{X,j}^\top \mathbf X+\epsilon_{M,j}, \\
  Y &= \beta_{0, j} + \beta_j M_j + \beta_{A,j}A+\boldsymbol\beta_{X,j}^\top \mathbf X+\epsilon_{Y,j},
\end{aligned}
\end{equation}
where $\epsilon_{M,j}$ and $\epsilon_{Y,j}$ are independent mean-zero error terms.  Model (\ref{Eq: mediator_models}) targets one mediator at a time, an approach that is advocated for large $J$ in recent papers \citep{liu2022large,Dai_2022,yang2025causal}.  Alternatively, one could jointly model the mediation effect by including all mediators in the outcome model.  For large $J$, inference requires advanced debiasing techniques for high-dimensional linear models \citep{derkach2020group, tian2022coxmkf}.  We focus on Model (\ref{Eq: mediator_models}) as p-values can be readily obtained by practitioners, and their goal in mediation testing is often to identify whether a potential variable is a mediator without considering its joint effect with other mediators.

Under Model (\ref{Eq: mediator_models}), a mediation relationship through $M_j$ requires evidence for both paths, i.e. $\delta_j\neq 0$ and $\beta_j\neq 0$, which naturally yields a composite hypothesis for each $j$:
$$
H_{0j}: \delta_j \beta_j = 0 \qquad \text{vs} \qquad H_{1j}: \delta_j \beta_j \neq 0.
$$
The composite null is true if either of the following base pathway null hypotheses is true:
\begin{equation}
\label{Eq: base_nulls}
\begin{aligned}
    H_{0\delta, j}: \delta_j = 0 \qquad \text{vs} \qquad H_{1\delta,  j}: \delta_j \neq 0 \ , \\
    H_{0\beta, j}: \beta_j = 0 \qquad \text{vs} \qquad H_{1\beta, j}: \beta_j \neq 0 \ .
\end{aligned}
\end{equation}
Hence $H_{0j}$ can be equivalently expressed as the union of the three combinations of the pathway hypotheses, namely
\begin{equation}
\label{Eq: null_cases}
\begin{aligned}
    & \text{Case I: }\ H_{0\delta, j}\ \text{ and } H_{1\beta, j}\ \text{ (partial null)}, \\
    & \text{Case II: }\ H_{1\delta, j}\ \text{ and } H_{0\beta, j}\ \text{ (partial null)}, \\
    &\text{Case III: }\ H_{0\delta, j}\ \text{ and } H_{0\beta, j}\ \text{ (global null)}.
\end{aligned}
\end{equation}

Traditionally, the Sobel test (\citealp{sobel1982asymptotic}) and joint significance test (\citealp{mackinnon2002comparison}) are common approaches to assess mediation effects for single or low-dimensional mediators. Sobel’s test formulates the test statistic as a product of two estimated regression coefficients $\hat\delta_j\hat\beta_j$, and uses a normal approximation based on delta method to construct a Wald-type standardized statistic for that product. In contrast, the joint significance test utilizes the asymptotically independent pathway-specific $p$-values $p_{\delta j}$ and $p_{\beta j}$, corresponding to the pathway hypotheses $H_{0\delta, j}$ and $H_{0\beta, j}$, and rejects the composite null $H_{0j}$ if $\max(p_{\delta j},p_{\beta j})$ is small, hence is also called the ``MaxP'' test. However, both Sobel test and MaxP test are conservative under the global null when $(\delta_j,\beta_j) = (0,\ 0)$ (\citealp{liu2022large}). For Sobel's test, the product map $(\delta_j, \beta_j) \mapsto \delta_j\beta_j$ is nonregular under global null, so the usual first-order delta-method normal approximation is invalid and the Sobel test statistic is asymptotically conservative.  Likewise, the MaxP test is also conservative under the global null since both $p_{\delta, j}$ and $p_{\beta, j}$ are uniform in this case, making $\mathbb P(\max(p_{\delta},\ p_\beta)\leq\alpha) = \alpha^2 \ll \alpha$ \citep{liu2022large}. 
Consequently, applying FDR-controlling procedures to Sobel and MaxP tests directly often results in very few discoveries in large-scale mediation studies, especially when the signals along one or both pathways are not particularly strong.

Recent advances, including DACT (\citealp{liu2022large}), HDMT (\citealp{Dai_2022}), and M-DACT (\citealp{yang2025causal}), addressing the low-power problem by explicitly estimating the proportions of the different null cases in (\ref{Eq: null_cases}).  These methods differ from each other in both how the proportions are estimated and % whether to apply the weights to directly adjust p-values or to change the rejection threshold of FDR-controlling methods.  
how the resulting estimates are incorporated into the testing procedure. DACT uses the estimated proportions as weights to construct composite $p$-values, HDMT uses them to estimate the mixture null distribution of the MaxP statistics and to adjust the rejection threshold, and M-DACT uses them for both weighting and threshold adjustment. After the adjustments, all three procedures apply an FDR controlling method to all $J$ hypotheses.  

In large-scale settings, even a well-calibrated FDR-controlling procedure must still pay a multiplicity penalty over all $J$ hypotheses, which can keep the rejection threshold stringent and limit sensitivity to weaker but scientifically meaningful mediators. Motivated by this limitation, we consider a two-stage procedure, inspired by the single-study two-stage framework of \cite{sarkar_2013}, which aims to further improve power by reducing the number of hypotheses that enter the final multiple-testing step. %Our goal is to increase testing power while maintaining control over the overall FDR rate. 
The general idea is straightforward: we first screen over all hypotheses and filter out the ones that are likely to be global null, then carry the remaining candidates forward to the testing stage. Usually, in large-scale setting with sparse signals, e.g. epigenome-wide mediation analysis or transcriptome-wide mediation analysis, this screening step could filter out most irrelevant DNA methylations or gene expressions, leaving us a much smaller subset for formal testing. Consequently, multiplicity adjustment in the second stage is applied to far fewer than $J$ hypotheses, yielding much less stringent rejection thresholds and will likely increase the power of detecting true signals among the retained candidates.

A key challenge in two-stage procedures is that the screening and testing p-values are generally dependent. Although the framework in \cite{sarkar_2013} is general enough to accommodate such dependence, a rather complicated adjustment in the second stage is needed.  In particular, this adjustment depends on the global null distribution (or copula) of dependent p-values, which is typically difficult to obtain in practice.  A major contribution of this work is to propose a two-stage procedure that capitalizes on a special structure of testing high-dimensional mediation hypotheses, so that the first-stage screening p-value is asymptotically independent of the second-stage testing p-value. 
The first stage is designed to screen out the more stringent global nulls in (\ref{Eq: null_cases}), which often constitute the majority of the hypotheses in large-scale testing scenarios. FDR-controlling methods are then applied to the remaining hypotheses in the second stage.   
Although both stages use the same full sample, the resulting second-stage p-value is asymptotically independent of the first-stage screening p-value.
The key insight is that the screening and testing p-values are transformations of the order statistics of the two pathway-specific p-values, and the asymptotic independence between the transformations can be established using R{\'e}nyi's classical representation of ordered standard uniform distribution \citep{renyi1953theory}. This allows the same full data to be used in both stages without sample splitting.  We then show that the proposed two-stage procedure controls FDR asymptotically. \cite{derkach2020group} recently proposed a multi-step procedure that shares the high-level intuition of our proposed two-stage approach, but requires user-defined or pre-determined screening thresholds and further adjustments in subsequent steps. In contrast, our proposed method adopts a data-adaptive screening threshold that adapts to the unknown signal sparsity and strength, and does not require adjustment in the testing stage due to asymptotic independence.

We will describe the method in detail in Section \ref{sec: method}, including the construction of transformed p-values, the adaptive screening algorithm in the first stage, and the FDR-controlled testing procedure used in the second stage. The same section also establishes the main theoretical properties, including asymptotic independence of the transformed $p$-values and asymptotic FDR control of the proposed procedure. Section \ref{sec: sim} reports results from simulation studies comparing proposed method with DACT, HDMT, and M-DACT across a range of signal sparsity and signal strength settings. The proposed procedure will be applied to the Normative Aging Study to further evaluate its performance relative to the competing methods in Section \ref{sec: data}.  Final remarks are given in Section \ref{sec: disc}.

\section{Methods}
\label{sec: method}

Suppose we have $J$ potential mediators, i.e. $J$ mediation hypotheses to be tested. For the $j$-th mediator, we have a pair of $p$-values, $(p_{1j}, p_{2j})$, corresponding to the path-specific hypotheses $H_{01, j}$ and $H_{02, j}$ in (\ref{Eq: base_nulls}). In large-scale mediation testing, $J$ is usually very large, typically on the order of tens or even hundreds of thousands, so traditional multiple testing adjustments are overly conservative. In such cases, a preliminary screening could be very effective in eliminating candidates that are highly likely to be null, e.g. candidates with large $\min(p_{1j}, p_{2j})$, and hence greatly increase the testing power.  However, screening must be handled with care: if the screening statistic is dependent on the downstream testing statistics under the null, then conditioning on “passing the screen” can distort the null distribution of the test statistics and lead to a failure to control on error rates (\citealp{bourgon2010independent}). As a result, valid two-stage procedures often require either (i) screening criteria that are independent of the testing $p$-values, which is done, for instance, using sample splitting, or (ii) additional adjustments on testing statistics to account for dependency, which impairs the testing power and complicates the development of FDR control method (\citealp{sarkar_2013}). These observations motivate our usage of a pair of $p$-values that (1) preserve the information of ordered p-values needed for effective screening, and (2) being asymptotically independent across the screening and testing stages, so that standard FDR procedures can be applied in the second stage without requiring further adjustment for selection-induced dependence.

Let $p_{(1)j} \leq  p_{(2)j}$ be the ordered base p-values for the $j$-th hypothesis, where the pair of base p-values, $(p_{1j},p_{2j})$ for testing $H_{0\delta,j}$ and $H_{0\beta,j}$, are asymptotically independent due to score orthogonality \citep{liu2022large,Dai_2022}.  We consider the p-value transformations:
$$p^*_{1/2, j} := 1 - \left[ 1 - p_{(1)j} \right]^{2},\quad p_{2/2, j}^* = 1-\left[\frac{1-p_{(2)j}}{1-p_{(1)j}}\right].$$
Here, $p^*_{1/2, j}$ is the transformed minimum p-value of \cite{tippett1931methods} derived directly from the distribution of the minimum of two uniform random variables, and we call $p^*_{2/2, j}$ a rescaled $p$-value. By definition, $p^*_{1/2, j}$ is small if $p_{(1)j}$ is small, and the relationship between $p_{(1)j}$ and $p^*_{1/2, j}$ is monotone but nonlinear.   In general, $p^*_{1/2, j}$ tends to be at least at large as $p_{(1)j}$, so using $p^*_{1/2, j}$ for stage-1 filtering yields a more stringent screening rule than using $p_{(1)j}$ is used for filtering directly. As a result, fewer hypotheses are carried forward to Stage 2, which further reduces the number of hypotheses entering the final multiple-testing adjustment and may in turn leads to less stringent rejecting threshold in Stage 2. 

The relationship between $p_{(2)j}$ and $p^*_{2/2, j}$, on the other hand, is neither linear nor monotone, with $p^*_{2/2, j}$ is generally not larger than $p_{(2)j}$, making the rejection rule more lenient than using $p_{(2)j}$ directly. 
Among hypotheses that passed stage 1, $p_{(1)j}$ must be small, hence 
%a small gap implies that $p_{(2)j}$ is also small. Consequently, although the relationship between $p_{(2)j}$ and $p^*_{2/2, j}$ is not monotone marginally, 
a small $p^*_{2/2, j}$ still corresponds to a small $p_{(2)j}$ conditional on passing the filtering step. In this sense, the rescaled $p$-values retain the signal-ordering information in the original ordered $p$-values while gaining additional desirable properties for the two-stage procedure, stated as follows. %According to \cite{TR-P}, we have the following proposition:

\begin{proposition}
     If $(p_{1j}, p_{2j})$ satisfies the assumption that $(p_{1j}, p_{2j})\stackrel d\to (D_1, D_2)$ as sample size $n\to\infty$, where $(D_1, D_2)$ are mutually independent, and for $k = 1, 2$, $D_k$ is standard uniform if $H_{0k, j}$ is true, and is a degenerating random variable with point mass at 0 if $H_{0k, j}$ is false, then the transformed $p$-values $p_{1/2}^*$ and $p_{2/2}^*$ satisfy the following properties:
     \begin{enumerate}
    \item Under $H_{0j}$, $p^*_{2/2, j} \stackrel d \to Unif(0, 1)$.
    \item Under $H_{1j}$, $p^*_{2/2, j} \stackrel p \to 0$.
    \item Under $H_{0j}$ or $H_{1j}$, $ p^*_{1/2, j}$ and $p^*_{2/2, j}$ are asymptotically independent.
\end{enumerate}
\label{prop: TR-P}
\end{proposition}
The proof is provided in the Supplementary Materials. 

With this pair of asymptotically independent screening and testing $p$-values and a prefixed FDR level $\alpha$, a general outline of the two-stage procedure is stated in Algorithm \ref{alg: two-stage}. The goal of Stage 1 is to filter out hypotheses with both $H_{0\delta, j}$ and $\ H_{0\beta, j}$ being null, \emph{i.e.} case 3 in (\ref{Eq: null_cases}), which usually make up a large proportion of potential mediators in genomics applications.  The screening thus greatly reduces the number of hypotheses being tested in Stage 2, subsequently increases testing power through a less stringent cost of multiple comparison. 

\begin{algorithm}
\caption{Two-stage procedure}
\label{alg: two-stage}
\begin{enumerate}
  \setlength{\itemsep}{0.2em}
  \item[] \textbf{Input}: Stage 1 threshold $t_\gamma$, desired FDR level $\alpha$, transformed p-values $p_{1/2,j}^*, p_{2/2,j}^*$ for $j=1,\ldots,J$.
  \item[] \textbf{Stage 1 (Screening)}: Let $I(t_\gamma) = \{j: p_{1/2,j}^*\leq t_\gamma\}$ denotes the index set of hypotheses passed Stage 1. Let $R_1(t_\gamma) = |I(t_\gamma)|$ denote the number of hypotheses passed Stage 1, where $0\leq R_1(t_\gamma)\leq J$. If $R_1(t_\gamma)=0$, reject no hypothesis and stop; otherwise, proceed to Stage 2. 
    \item[] \textbf{Stage 2 (Selection)}: Let $t_\alpha = \sup\left\{t: \widehat {FDR}\ (t; t_\gamma)\leq\alpha\right\}$ denote the Stage 2 threshold, where $\widehat {FDR}\ (t; t_\gamma)$ can be any general estimator of FDR. We reject the composite hypothesis $H_{0j}$ if $j\in I(t_\gamma)$ and $p_{2/2, j}^*\leq t_\alpha$.
\end{enumerate}
\end{algorithm}

The selection of $t_\gamma$ plays an important role here as it cannot be too large or too small. A large $t_\gamma$ allows many hypotheses to pass through Stage 1, but this inflates the multiple-testing burden in Stage 2, requiring a more stringent rejection threshold, which may reduce statistical power. Meanwhile, a small $t_\gamma$ may prematurely filter out true nonnull hypotheses in Stage 1, also resulting in a loss of power. Therefore, an ideal choice of $t_\gamma$ should preserve as much potential nonnulls as possible, while being able to screen out most global nulls.
With these considerations in mind, we adopted the signal missing rate (SMR) framework introduced by \cite{jeng2019efficient} for selecting Stage 1 threshold. For any $\epsilon\in (0, 1)$, the signal missing rate is defined as $SMR^\epsilon =\mathbb P(FN/s > \epsilon)$, where $FN$ is the number of false negatives and $s$ is the total number of true signals. In other words, the signal missing rate measures the probability of ignoring more than an $\epsilon$ proportion of the true signal. \cite{jeng2019efficient} proposed two procedures, cvSMR and adSMR. The former controls SMR at a prespecified level, however, it is conservative in terms of false-negative control under high dimensionality, often retaining too many hypotheses and thereby increasing the burden on the downstream testing step. In contrast, adSMR is less conservative.  Using a consistent estimator of the total number of signals $s$, it achieves $SMR^\epsilon \to0$ while typically selecting a smaller set of hypotheses and introducing fewer false positives. Given our goal of retaining as much signal and as little noise as possible after Stage 1, we adopted the adSMR procedure to determine the Stage 1 threshold $t_\gamma$. The resulting procedure is detailed in Algorithm \ref{alg: SMR}, which modifies Algorithm 1 in \cite{jeng2019efficient} to match our setting. Following \cite{jeng2019efficient}, we use the estimator of \cite{meinshausen2006estimating} as the default estimator of the signal proportion, since it is consistent under block dependence and is applicable across both sparse- and dense-signal regimes. The framework also allows alternative proportion estimators to be substituted, and further discussion is given in Section \ref{sec: disc}.  %  \textcolor{blue}{[description of SMR and procedure in Jeng]}.

\begin{algorithm}
\caption{Selecting $t_\gamma$ using AdSMR with Meinshausen-Rice (MR) estimator}
\label{alg: SMR}
\begin{enumerate}
  \setlength{\itemsep}{0.2em}
  % \item  Simulate the bounding sequence $c_J$ from the empirical null distribution of $V_J$ with $\alpha_J = 1/\sqrt{\log J}$. 
  \item Obtain the estimated proportion of nonnulls using the estimator proposed by Meinshausen and Rice (2006): $\hat{\pi}_{\text{MR}} = \sup_{t\in(0,1)}\frac{F_J(t)-t-c_J\sqrt{t(1-t)}}{1-t}$, where $c_J$ is a bounding sequence and $F_J(t)$ is the empirical distribution of filtering p-values $p^*_{1/2}$.
  \item Sort the filtering $p$-values as $p^*_{1/2,\ (1)} \leq p^*_{1/2,\ (2)} \leq \cdots \leq p^*_{1/2,\ (J)}$. 
  \item Calculate the cut-off position as $k^*=\hat s + \min\left\{j\geq 1:p^*_{1/2,\ (\hat s + j)}\leq \frac{j}{J - \hat{s}}\right\}\mathbb I(\hat s >t_1)$, where $\hat{s} = J\hat{\pi}_{\text{MR}}$ and $t_1 = \max\{j: p^*_{1/2,\ (j)}<\frac{1}{J\sqrt{\log J}}\}$. Set an upper limit for $k^*$ at $\lfloor J/2 \rfloor$.
  \item Set $t_\gamma = p_{1/2,\ (k^*)}^*$.
\end{enumerate}
\end{algorithm}

By default, the adSMR procedure identifies the bounding sequence $c_J$ via simulation to satisfy $\mathbb P(V_J>c_J)=1/\sqrt{\log J}\to 0$, where $V_J=\sup_{t\in(0, 1)} \frac{U_J(t)-t}{\sqrt{t(1-t)}}$ and $U_J$ denotes the empirical distributions of $p$-values when all $J$ hypotheses are null. The simulation-based calibration was proposed to account for the block dependence structure among hypotheses commonly encountered in epigenome-wide association studies. The corresponding $c_J$ is denoted by $c_J^{sim}$.  It performs well with reasonable computational cost for moderate $J$. However, when the number of hypotheses grows to hundreds of thousands, the simulation process becomes computationally intensive. In such cases, the bounding sequence $c_J$ can be replaced by the closed-form Gumbel approximation from \cite{meinshausen2006estimating}, denoted by $c_J^{ind}$, which is derived under an independence assumption but avoids simulation entirely. A sensitivity analysis in the Supplementary Materials confirms that the resulting $\hat\pi_{MR}$ using $c_J^{ind}$ remains close to that obtained via simulation ($c_J^{sim}$) in the presence of dependence, and the discrepancy diminishes as $J$ increases due to the improved stability of the estimator at larger number of hypotheses. Since we observe $c_J^{ind} \leq c_J^{sim}$ consistently under positive dependence, the independence-approximated bounding sequence consistently produces a slightly smaller estimated null proportion, which means the resulting screening threshold will allow more hypotheses retained in Stage two. While this would be problematic if the bounding sequence were directly responsible for error rate control, in our procedure the FDR or FWER guarantee is provided solely by the Stage-2 testing procedure and holds for any data-adaptive choice of Stage 1 screening threshold $t_\gamma$ due to martingale properties of Stage-2 procedures \citep{STS_2004} with an extended filtration including Stage-1 p-values. Consequently, allowing slightly more hypotheses into Stage 2 does not compromise the overall FDR control, and may in fact slightly improve power by reducing the chance of screening out true signals.

In general, any p-value-based screening methods applied to the transformed p-values $p^*_{1/2, j}$ can be used in Stage 1. There is no universal best choice as the effectiveness of each approach varies across scenarios and should be decided accordingly.  An alternative Stage 1 threshold selection method is given in the Supplementary Materials.

In Stage 2, any step-up FDR-controlling method can be applied. We focus on step-up procedures in this work because they tend to be more powerful than their step-down counterparts (\citealp{BKY_2006}; \citealp{gavrilov2009adaptive}). Here, our default choice of FDR estimator is by \cite{STS_2004}:
$$\widehat{FDR}_\lambda\ (t; t_\gamma) = \begin{cases} \frac{\hat\pi_0(\lambda; t_\gamma)\cdot t}{\{R_2(t; t_\gamma)\vee 1\}/R1(t_\gamma)} & \text{if } t\leq\lambda \\ 1 & \text{if } t>\lambda\end{cases}$$
for a prespecified $\lambda > 0$, where 
$$\hat\pi_0\ (\lambda;t_\gamma) = \frac{R_1(t_\gamma)-R_2(\lambda; t_\gamma)+1}{(1-\lambda)R_1(t_\gamma)}$$
is the Storey's estimator of the proportion of true nulls out of the $R_1(t_\gamma)$ hypotheses in Stage 2, and $R_2(\lambda; t_\gamma) = \#\{j:p_{2/2, j}^*\leq \lambda,\ j\in I(t_\gamma)\}$.
The following theorem states that the two-stage procedure can control the FDR at a prespecified level asymptotically with this default estimator under minimal assumptions. 
\begin{theorem}[FDR control for Two-stage Procedure With Storey's FDR estimator]
\label{thm: fdr_storey}
Under the assumption that the null $p$-values are mutually independent asymptotically, with a prefixed $\alpha\in(0,1)$, the proposed two-stage procedure with default Storey's estimator in Stage 2 controls the false discovery rate at level $\alpha$ asymptotically.
\end{theorem}
Proof of Theorem \ref{thm: fdr_storey} is provided in the Supplementary Materials. 
\begin{remark}
\label{rmk: fdr_storey}
Under a stronger assumption that all $p$-values are mutually independent, Theorem \ref{thm: fdr_storey} can be extended to allow any general linear step-up procedure in Stage 2, provided its estimator of the proportion of true null hypotheses is conservative and coordinate-wise nondecreasing in the Stage 2 $p$-values.  The proof is provided in the Supplementary Materials.
\end{remark}
While the theoretical results rely on mutual independence across $j$, standard step-up FDR procedures are known to be robust to positive regression dependency.  The empirical performance under block dependence is validated using simulations in Section \ref{sec: sim}.

The proposed procedure uses two different proportion estimators by default, with the estimator of \cite{meinshausen2006estimating} in the screening stage and the estimator of \cite{STS_2004} in the testing stage. This design choice reflects the distinct statistical environments encountered at each stage. In Stage 1, we usually have a large number of hypotheses with sparse signals. 
In such extremely sparse settings, the Meinshausen-Rice estimator is favored because it is better at detecting a small signal proportion.  In the Normative Aging Study in Section \ref{sec: data}, for example, Meinshausen-Rice yields $\hat\pi_{00}=0.9993$
while Storey's estimator gives $\hat\pi_{00} = 1$, which would fail to detect any signal. After Stage 1, most global nulls are screened out, which substantially increases the proportion of nonnulls and reduces signal sparsity in Stage 2. Therefore, estimators designed for relatively dense signal proportions, such as Storey's estimator, become preferable.

Asymptotically, the FDR control of the proposed procedure is determined entirely by the Stage-2 procedure and is unaffected by the choice of screening threshold used in Stage 1. This advantage follows from the asymptotic independence between the Stage-1 filtering $p$-values and the Stage-2 testing $p$-values, which prevents the filtering step from distorting the null distribution of the test inputs in Stage 2. This also means that the FDR-controlling procedure used in the second stage can be replaced by any procedure with a known error-rate guarantee applied to independent p-values to control a different error rate metric. For example, replacing it with an adaptive Bonferroni correction would yield control on the family-wise error rate (FWER) instead of FDR, without requiring any modification of the Stage 1 procedure.

\section{Simulation}
\label{sec: sim}

%Our simulation design closely follow the settings in \cite{liu2022large}. 
In the first set of simulations, we followed the data generation procedure as in \cite{liu2022large}.  In particular, the exposure $A$ was generated from the Bernoulli distribution with a success rate of 0.5; two covariates, $X_1\sim N(10,\ 1)$ and $X_2\sim N(5,\ 1)$, were also generated.  We considered four sets of alternative $(\delta_a,\ \beta_a)$ configurations: $(0.3,\ 0.133)$, $(0.2,\ 0.2)$, $(0.133,\ -0.3)$, $(0.1,\ 0.4)$, which cover a range of effect magnitudes and directions for the underlying mediation components. To reflect varying degrees of signal sparsity, we examined four combinations of $(\pi_{00},\ \pi_{11})$, where $\pi_{00}$ denotes the proportion of composite hypotheses with both base hypotheses being null, and $\pi_{11}$ is the proportion of nonnull composite hypotheses. Specifically, we included one sparse case $(0.98,\ 0.01)$, one moderate case $(0.9, 0.04)$, and two dense cases $(0.8,\ 0.1)$ and $(0.4, 0.2)$. We set $\pi_{10}$ and $\pi_{01}$, the proportions of hypotheses with partial nulls, to be $\pi_{10}=\pi_{01}=(1-\pi_{00}-\pi_{11})/2$. For global nulls, we set the true parameter values $(\delta_j,\ \beta_j)$ to be $(0,\ 0)$; for partial nulls, we set one of $(\delta_j,\ \beta_j)$ to be 0, and the other to be the corresponding alternative value. For example, under the scenario with $(\delta_a,\ \beta_a)=(0.3,\ 0.133)$, hypotheses belong to the $\pi_{10}$ proportion will have $(\delta_j,\ \beta_j)=(0.3,\ 0)$. Similarly, hypotheses belong to the $\pi_{01}$ proportion will have $(\delta_j,\ \beta_j)=(0,\ 0.133)$.  The mediator $M_j$ was generated as
$$
M_j = \delta_jA + 0.2X_1 + 0.3 X_2 + \epsilon_M,
$$
where $\epsilon_M\sim N(0,\ 1)$, and the outcome $Y_j$ were generated as 
$$
Y_j = A + \beta_j M_j + 0.1 X_1 + 0.2X_2 + \epsilon_Y,
$$
with $\epsilon_Y\sim N(0,\ 2)$. 
The sample size was varied across $n\in\{800,\ 1000,\ 1200\}$, and the number of hypotheses was set to $J=10,000$ to reflect the large-scale context.

\begin{figure}[!p]
  \centering
  \includegraphics[width=1\textwidth, alt = {A two-by-two grid of Q-Q plots showing observed p-values on the negative log base 10 scale (vertical axis) against expected null p-values (horizontal axis), for the parameter settings delta and beta equal to (0.1, 0.4), (0.133, negative 0.3), (0.2, 0.2), and (0.3, 0.133). In every panel, the rescaled p-values (blue) lie along the dashed 45-degree reference line, indicating correct null calibration, while the original p-values (red) fall consistently below the line, showing they are systematically conservative.}]{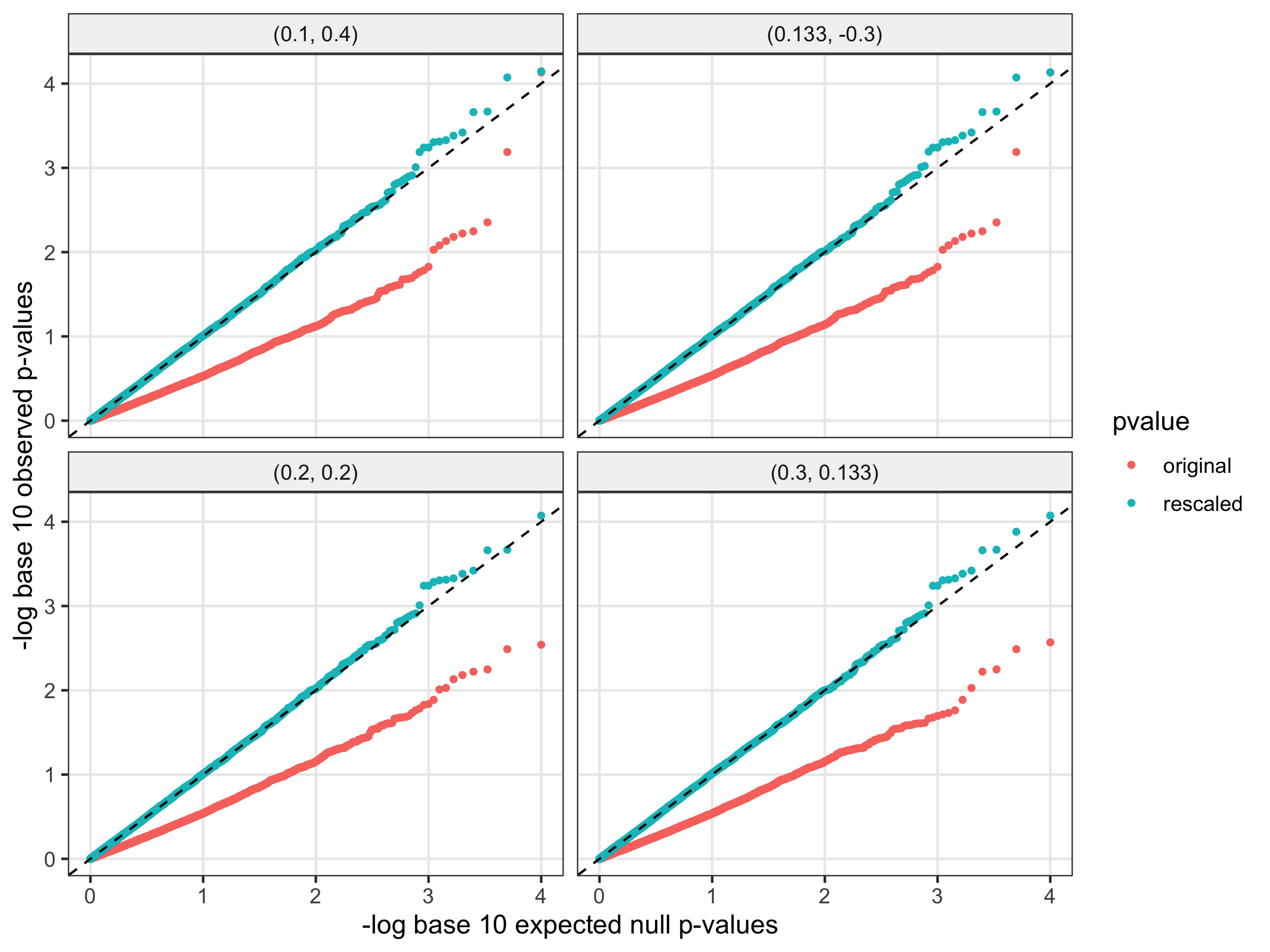}
  \caption{Q-Q plot comparing the original ordered p-values $p_{(2)j}$ and the rescaled p-values $p_{2/2, j}^*$, with $(\pi_{00},\pi_{11})=(0.9, 0)$, and $\pi_{10} = \pi_{01} = 0.05$. Each panel shows ordered observed p-values against the corresponding expected p-values generated from $Unif(0,1)$ reference distribution on the $-\log_{10}$ scale; the dashed 45$^\circ$ line indicates exact null calibration. Panels correspond to the four parameter settings $(\delta,\beta)=(0.1,0.4),\ (0.133,-0.3),\ (0.2,0.2),\ (0.3,0.133)$.}
  \label{fig: sim_QQplot}
\end{figure}

For each scenario, we fit a linear regression of \texttt{$M_j \sim A + X_1 + X_2$} to obtain $p_{1j}$ corresponding to the base hypotheses $H_{0\delta,\ j}: \delta_j= 0$, and fit the regression \texttt{$Y_j \sim A + M_j+ X_1 + X_2$} for the base hypotheses $H_{0\beta,\ j}: \beta_j=0$ to obtain $p_{2j}$.  %We then apply the transformation introduced in \cite{Yuan_2024} to obtain the rescaled $p$-values $p^*_{1/2, j}$ and $p^*_{2/2, j}$, which we will be using to apply the proposed method. 
Figure \ref{fig: sim_QQplot} presents Q-Q plots comparing the observed original ordered p-values $p_{(2)j}$ and the rescaled p-values $p_{2/2, j}^*$ against the expected null distribution, each using one dataset generated under $n=1000$. In all scenarios, the original ordered quantities deviate substantially from the null reference, whereas the rescaled stage-2 $p$-values lie much closer to the diagonal.

% fig 1

\begin{figure}[!p]
  \centering
  \includegraphics[width=1\textwidth, alt = {A three-by-four grid of bar charts comparing five mediation testing procedures - DACT, HDMT asymptotic, HDMT empirical, M-DACT, and the Proposed method - across four combinations of the null proportions pi-zero-zero and pi-one-one. Rows show empirical power (top), empirical false discovery rate with a dashed line at the nominal level of 0.20 (middle), and empirical family-wise error rate with a dashed line at the nominal level of 0.05 (bottom). DACT attains the highest power but grossly violates both error rates, with false discovery rate roughly 0.7 to 0.85 and family-wise error rate up to about 0.78. The other four methods control both error rates at their nominal levels, and among these valid methods the proposed method attains the highest power.}]{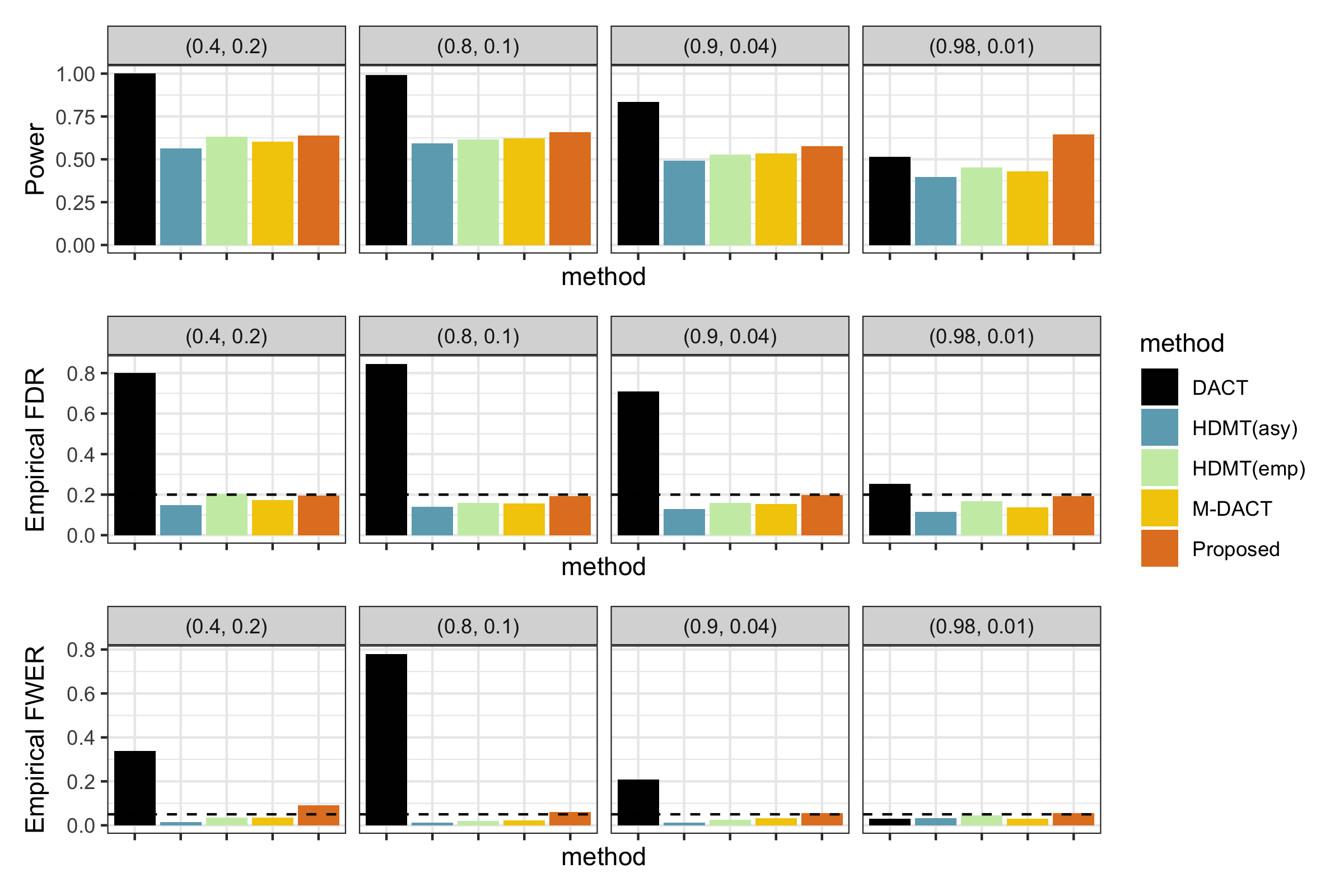}
  \caption{Empirical power (top row), FDR (middle row), and FWER (bottom row) of five mediation testing procedures for each $(\pi_{00}$,\ $\pi_{11})$ combination. Results are averaged across the remaining simulation configurations, including effect sizes $(\delta,\beta)$ and sample size $n$. The top two rows evaluate procedures controlling FDR at $\alpha = 0.2$ (dashed line), while the bottom row evaluates procedures controlling FWER at $\alpha = 0.05$ (dashed line).}
  \label{fig: sim_result}
\end{figure}

We compared the proposed method with HDMT(asy), HDMT(emp), DACT, and M-DACT. For each method, we evaluated empirical power and error rates across simulation replicates. The upper row of Figure~\ref{fig: sim_result} summarizes the simulation results with the target FDR level fixed at $\alpha = 0.2$. Each panel was averaged across sample sizes and effect sizes with 100 replications for each configuration, resulting in 1200 independently generated datasets per panel.  When $(\pi_{00}, \pi_{11}) = (0.98,\ 0.01)$, the proposed procedure outperforms DACT, M-DACT, and both versions of HDMT. In the remaining scenarios, it performs comparably to, and often slightly better than, HDMT and M-DACT. Across all examined scenarios, the proposed method attains accurate FDR control at the nominal level, while DACT exhibits inflated FDR and fails to maintain control in the settings examined. HDMT, in contrast, is generally conservative, with empirical FDR falling below $\alpha$ in most scenarios. 

% \textcolor{blue}{[Add description for FWER sim results]}

We further evaluated all methods under FWER control at $\alpha = 0.05$, using the same simulation configurations as in the FDR study. Our proposed procedure readily accommodates FWER control by applying an adaptive Bonferroni correction at Stage 2 in place of the BH procedure. To extend DACT to the FWER setting, we similarly replaced the BH adjustment with an adaptive Bonferroni correction. HDMT(asy), HDMT(emp), and M-DACT were applied with their built-in FWER-controlling options. As shown in the bottom row of Figure \ref{fig: sim_result}, HDMT(asy) shows conservative FWER across all configurations, while DACT exhibits slightly conservative FWER when $(\pi_{00},\ \pi_{11}) = (0.98,\ 0.01)$ and shows substantial inflation in all other settings. The remaining methods all control FWER below or close to the 0.05 threshold across all configurations.

% Fig 2

The scenarios above treat the mediators $M_j$ as mutually independent. In practice, however, candidate mediators such as DNA methylation sites or gene-expression probes may be correlated within genomic regions or co-regulated pathways, and true mediators may be concentrated within a limited number of such regions rather than scattered uniformly across the genome. To assess robustness under more realistic dependence, we conducted an additional simulation study focusing on the sparse configuration $(\pi_{00},\ \pi_{11}) = (0.98,\ 0.01)$. We imposed a block-exchangeable correlation structure on the mediators, with block size varying over ${10,50,100}$ and within-block correlation fixed at $\rho=0.3$. 
The $\pi_{11}$ true mediators were concentrated within designated signal blocks, with an expected proportion of $0.8$ nonnull mediators in each signal block. Accordingly, the number of signal blocks was set to $\lceil M\pi_{11}/(\text{block size}\times \text{signal proportion})\rceil$. To accommodate the dependence among mediators, the outcome was generated from a joint model
$$
Y = A + \sum_{j:\ b_j\neq 0} \beta_j M_j + 0.1 X_1 + 0.2 X_2 + \epsilon_Y,
$$
in which all mediators with a nonzero $\beta$-path contribute simultaneously, in contrast to the per-mediator outcome used above. Because the testing procedures fit one mediator at a time, their population target is the marginal rather than joint $\beta$-path. Under the block-exchangeable covariance structure, the marginal coefficient for mediator $j$ is
$$
\beta_j^{\mathrm{marg}}=\beta_j+\rho\sum_{\substack{k\neq j,\ k\text{ in the same block}}}\beta_k,
$$
whereas the marginal exposure-mediator coefficient remains $\delta_j$. Consequently, a mediator with $\delta_j\neq0$ and $\beta_j=0$ in the joint model may have a nonzero marginal mediation effect when it shares a block with a mediator having $\beta_k\neq0$. We therefore evaluated FDR, FWER, and power using the marginal mediation effects, defining mediator $j$ as nonnull when $\delta_j\beta_j^{\mathrm{marg}}\neq0$, rather than using its original generative category. This raises the effective $\pi_{11}$ by roughly $0.0002$ to $0.002$ as the block size increases from 10 to 100. All remaining configurations, including effect sizes, sample sizes, the distribution of $A$, and the mediator model for $M_j$, were left unchanged. 

\begin{figure}[!p]
  \centering
  \includegraphics[width=1\textwidth, alt = {A three-by-three grid of bar charts for the same five methods under the sparse setting where the null proportions pi-zero-zero and pi-one-one equal (0.98, 0.01), with a block-exchangeable correlation structure of 0.3. Columns correspond to block sizes of 10, 50, and 100. Rows show empirical power (top), empirical false discovery rate with a dashed line at the nominal 0.20 (middle), and empirical family-wise error rate with a dashed line at the nominal 0.05 (bottom). The Proposed method has the highest power at every block size while holding the false discovery rate at the nominal 0.20; the family-wise error rate stays near or below 0.05 for all methods, with the Proposed method slightly exceeding it at block size 50 and HDMT(emp) slightly inflated at block size 100.}]{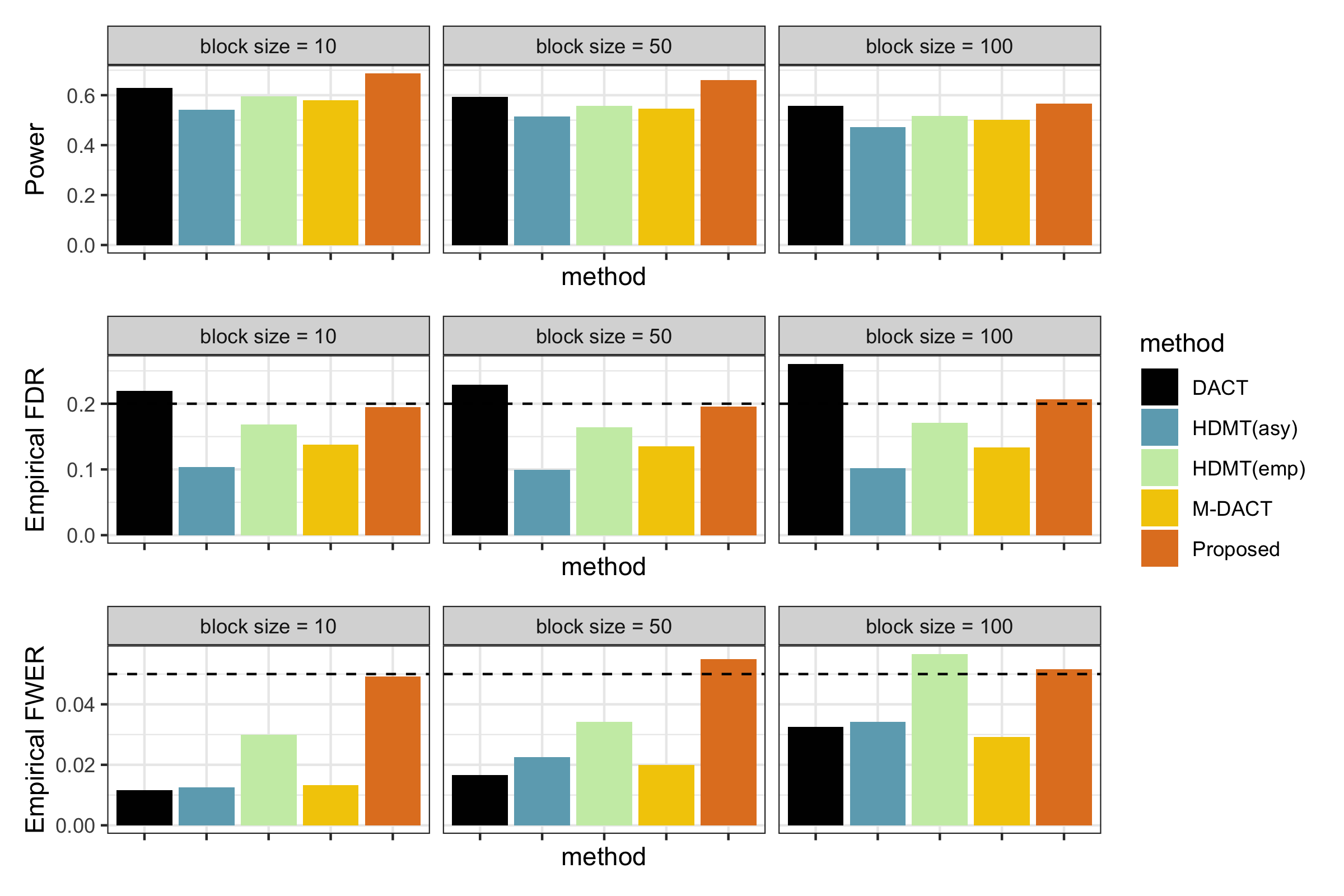}
  \caption{Empirical power (top row), FDR (middle row), and FWER (bottom row) of five mediation testing procedures under $(\pi_{00}, \pi_{11}) = (0.98, 0.01)$. Mediators follow a block-exchangeable correlation structure with within-block correlation $\rho=0.3$, and true mediators are concentrated within signal blocks with an expected within-block nonnull proportion of $0.8$. Results are averaged across the remaining simulation configurations, including effect sizes $(\delta,\beta)$ and sample size $n$. Columns correspond to block sizes of 10, 50, and 100. Dashed horizontal lines indicate the nominal FDR level of 0.20 and nominal FWER level of 0.05.}
  \label{fig: sim_corr}
\end{figure}

Figure \ref{fig: sim_corr} summarizes the results. Under FDR control, the empirical power of all five procedures was relatively stable across block sizes. The proposed procedure achieved the highest power throughout, with its advantage slightly narrowed when the block size was 100. Its empirical FDR remained close to and below the nominal level of 0.20 across block sizes. DACT, in contrast, exceeded the nominal FDR level for all three block sizes, with the degree of inflation increasing as the blocks became larger. The two HDMT procedures and M-DACT were generally conservative, particularly HDMT(asy). 
Under FWER control, the proposed procedure remained close to the nominal level across block sizes. HDMT(emp) was also close to $0.05$ at block size 100, although it was more conservative at block sizes 10 and 50. In contrast, DACT, M-DACT, and HDMT(asy) were consistently conservative across all settings. 

% Fig 3

\section{The Normative Aging Study}
\label{sec: data}

Smoking is a well-established cause of various lung diseases (\citealp{jayes2016smokehaz}), yet the molecular pathways linking the two are less well understood. DNA methylation offers one plausible route, as methylation levels have been reported to be associated with both lung function and smoking status (\citealp{joehanes2016epigenetic}; \citealp{lee2022pulmonary}). This motivates mediation analyses to identify DNA methylation CpG sites whose methylation changes mediate the association between smoking and lung function.

The Normative Aging Study (NAS) is a longitudinal study focusing on non-pathological aging, and followed a cohort of male participants for over 30 years (\citealp{bind2014air}). % Participants were screened for disease status at the time of entry into the study, and only subjects who were free of certain diseases were eligible, regardless of age at entry. 
Enrollment was open to men of any age, provided they were free of known chronic illness at entry. 
%(\citealp{bell1972normative}).
Participants were followed with repeated in-person examinations approximately every 3–5 years, during which the study collected detailed information on demographics, medical history, and lifestyle factors, along with standardized physical examinations and laboratory assessments, yielding a rich longitudinal dataset for studying determinants of health outcomes.

The NAS dataset consists of $n=603$ men with DNA methylation measurements at $J=484,613$ CpG sites (\citealp{liu2022large}). To identify DNA methylation CpG sites that potentially mediate the effect of smoking on lung functions, \cite{liu2022large} fitted the models in (\ref{Eq: mediator_models}), adjusting for age, height, weight, education and medication histories, estimated blood cell-type proportions, and five principal components that capture 95\% of the variation of DNA processing batch effects. In this analysis, $A$ is a binary exposure indicating smoking status, $M_j$ with $j = 1, \dots, J$ denotes the methylation level at the $j$th CpG site, and $Y$ is lung function measured as the forced expiratory flow over the mid-portion (25\%-75\%) of the Forced Expiratory Vital capacity (FEF$_{25\%-75\%}$). 

% Fig 4
\begin{figure}[!p]
  \centering
  \includegraphics[width=1\textwidth, alt = {Two sets of Q-Q plots of rescaled p-values from the NAS data application, plotted on the negative log base 10 scale against expected null p-values, with a dashed 45-degree line representing the Uniform null distribution. In panel (a), the screening p-values mostly follow the null line, while a group of points in the upper right rises above it and above the horizontal dotted screening threshold; these points pass the screening stage. In panel (b), results are split by Stage 1 outcome: the "Not selected in Stage 1" panel follows the null line with almost no points crossing the dotted rejection threshold, whereas the "Selected in Stage 1" panel shows a cluster of points rising above the line and above the threshold that are rejected in Stage 2.}]{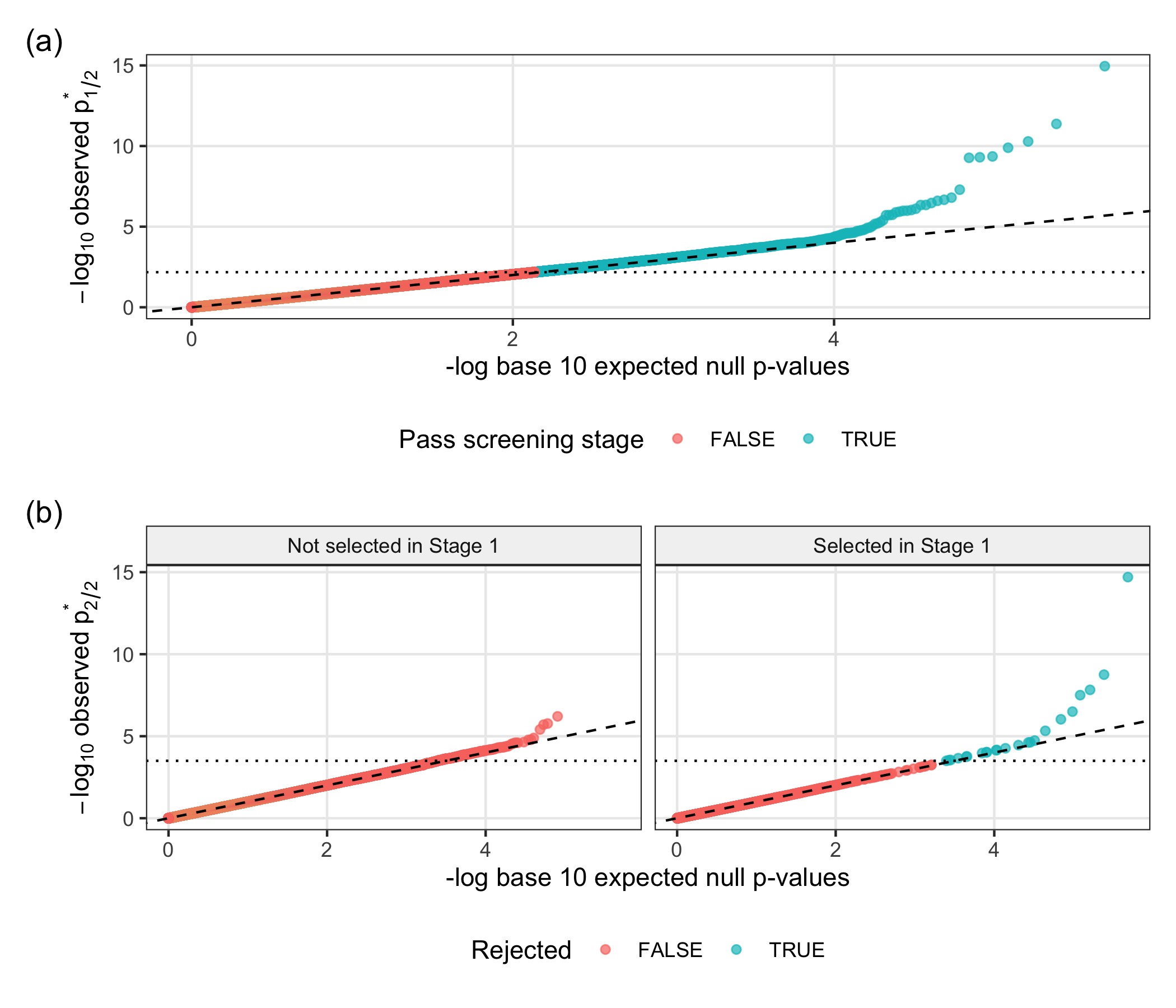}
  \caption{Q-Q plots comparing the rescaled p-values in the NAS data application against the expected null distribution. The dashed diagonal line represents the reference expected under a $\mathrm{Uniform}(0,1)$ null distribution.  (a) Observed screening p-values $p_{1/2}^*$ with points colored by whether they passed the screening stage; the horizontal dotted line marks the screening threshold. (b) Rescaled testing $p$-values $p_{2/2}^*$ stratified by Stage 1 screening outcome, with points colored by Stage 2 rejection status; the horizontal dotted line marks the rejection threshold.}
  \label{fig: realdat_QQplot}
\end{figure}

We reanalyzed the $p$-values provided by \cite{liu2022large}, which were obtained by fitting linear models (\ref{Eq: mediator_models}).
The estimated global null proportion is $99.933\%$ using the Meinshausen-Rice estimator with a simulated threshold suggested in \cite{jeng2019efficient}. Under the proposed two-stage method, 
3,557 hypotheses were retained after Stage 1, leaving us a much smaller subset of hypotheses to be tested. Importantly, the filtering step does not remove hypotheses with the smallest $p_{(2)j}$ values. As shown in Figure \ref{fig: realdat_QQplot} panel (a), the rescaled screening $p$-values $p_{1/2}^*$ align closely with the expected $\mathrm{Unif}(0,1)$ null distribution over most of the range, with the strongest departures at the upper tail on the $-\log_{10}$ scale.  The hypotheses retained after Stage 1 include the most significant signals in the tail of the distribution, indicating that the screening step primarily removes unpromising hypotheses while preserving the strongest candidates for further testing. Figure \ref{fig: realdat_QQplot} panel (b) displays the rescaled testing p-values $p_{2/2}^*$, stratified by screening outcome and colored by Stage 2 rejection result. Among hypotheses not selected in Stage 1 (left panel), the p-values remain close to the null line with only minor departures, whereas among those selected in Stage 1 (right panel), the most significant signals rise far above the null reference. In Stage 2, we employed Storey's FDR estimator to determine the rejection threshold $t_\alpha$. With FDR control level $\alpha = 0.05$, we identified 23 CpG sites. The list of identified CpG sites are presented in Table \ref{tab: CpG_sites}. For comparison, applying Stage 2 directly to all hypotheses without any filtering yields only 6 rejections, illustrating the power gained by the Stage 1 screening step, which sharply reduces the multiple-testing burden.

% Tab 1
\begin{table}[!p]
\centering
\caption{CpG sites discovered by the proposed method.}
\label{tab: CpG_sites}
\begin{tabular}{lccccccc}
\hline
CpG site & Sobel & MaxP & HDMT(asy) & DACT & HDMT(emp) & M-DACT & Proposed \\
\hline
cg05575921    &  $\checkmark$   &   $\checkmark$ &   $\checkmark$ &   $\checkmark$ &   $\checkmark$ &   $\checkmark$ &   $\checkmark$  \\
cg03636183    &     &   $\checkmark$ &   $\checkmark$ &   $\checkmark$ &   $\checkmark$ &   $\checkmark$ &   $\checkmark$  \\
cg06126421 &     &   $\checkmark$ &   $\checkmark$ &   $\checkmark$ &   $\checkmark$ &   $\checkmark$ &   $\checkmark$  \\
cg21566642 &     &   $\checkmark$ &   $\checkmark$ &   $\checkmark$ &   $\checkmark$ &   $\checkmark$ &   $\checkmark$  \\
cg05951221 &     &   $\checkmark$ &   $\checkmark$ &   $\checkmark$ &   $\checkmark$ &   $\checkmark$ &   $\checkmark$  \\
cg01940273 &     &    &   $\checkmark$ &   $\checkmark$ &   $\checkmark$ &   $\checkmark$ &   $\checkmark$  \\
cg14753356 &     &    &   $\checkmark$ &   $\checkmark$ &   $\checkmark$ &   $\checkmark$ &   $\checkmark$  \\
cg15342087 &     &    &   $\checkmark$ &   $\checkmark$ &   $\checkmark$ &   $\checkmark$ &   $\checkmark$ \\
cg21322436 &     &    &   $\checkmark$ &   $\checkmark$ &   $\checkmark$ &   $\checkmark$ &   $\checkmark$ \\
cg23771366 &     &    &   $\checkmark$ &   $\checkmark$ &   $\checkmark$ &   $\checkmark$ &   $\checkmark$ \\
cg23916896 &     &    &   $\checkmark$ &   $\checkmark$ &   $\checkmark$ &   $\checkmark$ &   $\checkmark$ \\
cg24859433 &     &    &   $\checkmark$ &   $\checkmark$ &   $\checkmark$ &   $\checkmark$ &   $\checkmark$ \\
cg25189904 &     &    &   $\checkmark$ &   $\checkmark$ &   $\checkmark$ &   $\checkmark$ &   $\checkmark$ \\
cg25949550 &     &    &   &   $\checkmark$ &   $\checkmark$ &   $\checkmark$ &   $\checkmark$ \\
cg06644428 &     &    &   &   $\checkmark$ &   $\checkmark$ &   $\checkmark$ &   $\checkmark$ \\
cg11660018 &     &    &   &   $\checkmark$ &   $\checkmark$ &   $\checkmark$ &   $\checkmark$ \\
cg11902777 &     &    &   &   $\checkmark$ &   $\checkmark$ &   $\checkmark$ &   $\checkmark$ \\
cg03991871 &     &    &   &   $\checkmark$ &   $\checkmark$ &   $\checkmark$ &   $\checkmark$ \\
cg14624207 &     &    &   &   $\checkmark$ &   $\checkmark$ &   $\checkmark$ &   $\checkmark$ \\
cg11554391 &     &    &   &    &   $\checkmark$ &   $\checkmark$ &   $\checkmark$ \\
cg21161138 &     &    &   &    &   $\checkmark$ &   $\checkmark$ &   $\checkmark$ \\
cg11279857 &     &    &   &    &   &    &   $\checkmark$ \\
cg23079012 &     &    &   &    &   &    &   $\checkmark$ \\
\hline
\end{tabular}
\end{table}

\cite{yang2025causal} summarizes the CpG mediators identified in the NAS analysis across several procedures, including the Sobel test, MaxP, HDMT(asy), HDMT(emp), DACT, and M-DACT. In Table \ref{tab: CpG_sites}, we append their results to the CpG sites found by our method.  The resulting discovery sets are nested, with the Sobel test being the most conservative (1 discovery), followed by MaxP (5 discoveries), HDMT(asy) (13 discoveries), DACT (19 discoveries), and HDMT(emp) and M-DACT, which tie for the largest set at 21 discoveries each. Applying our proposed method to the same set of $p$-values, we recovered all 21 CpG sites found by HDMT(emp) and M-DACT and identified two additional sites, cg11279857 and cg23079012. While cg11279857 has not previously been reported as a smoking–lung-function mediator to our knowledge, cg23079012 has been linked to both smoking status and lung function in prior work (\citealp{joehanes2016epigenetic}; \citealp{lee2022pulmonary}).

Because the NAS dataset involves a large number of candidate mediators, the simulation-based proportion estimator used above is computationally intensive.  
We therefore also applied the Meinshausen-Rice estimator under the independence assumption, which gave an estimated null proportion of 99.931\% and retained 3736 hypotheses for Stage 2 testing. This is slightly more than the 3,557 retained under the simulation-based version, which is expected. As observed in our simulations in the Supplementary Materials, the independent approximation tends to produce a slightly anti-conservative proportion estimate relative to the simulation-based estimator. Despite this difference in filtering, the Stage 2 rejections were identical to those in Table \ref{tab: CpG_sites}.

\section{Discussion}
\label{sec: disc}

In this article, we proposed a two-stage adaptive testing procedure for large-scale mediation hypotheses with asymptotic FDR control. Rather than combining the two pathway p-values into a single test statistic and calibrating it against an estimated composite-null distribution, our procedure transforms them into asymptotically independent p-values used in two sequential stages. The first stage screens out global nulls using a data-adaptive threshold, and the second stage applies an adaptive step-up procedure to the surviving hypotheses. The asymptotic independence between the two stages enables a conditioning argument for guaranteeing FDR control with a data-adaptive first-stage threshold.

Simulation studies demonstrate that the proposed procedure maintains reliable FDR control across a wide range of signal sparsity and strength settings. In contrast to DACT, which showed inflated FDR under relatively dense signals, the proposed method remained stable and achieved substantial power gains over competing methods, especially in sparse settings with large $\pi_{00}$, where the screening step effectively removed most uninformative hypotheses and reduced the multiplicity burden in Stage 2. The analysis of the Normative Aging Study further confirmed the simulation findings and highlighted the practical advantage of the method. The adaptive screening step retained only a small subset of hypotheses, consistent with the extreme sparsity of mediation signals in this dataset, and the second stage identified significant CpG sites that were consistent with the findings of previous analyses by \cite{liu2022large} and \cite{yang2025causal}, while containing additional discoveries.

Several extensions and refinements of the proposed framework deserve further exploration. First, while the current procedure adopts the Meinshausen-Rice estimator and the Storey estimator as default estimators for the screening and testing null proportions, other combinations of screening algorithms, proportion estimators, and test statistics may perform better in particular regimes and worth study. Additionally, the two-stage framework could be extended to other composite-null testing problems beyond mediation, such as replicability analysis \citep{heller2014replicability}, where pairs of $p$-values arise naturally and the null hypothesis has a similar composite structure. However, the validity of such extensions would require problem-specific justification, since the FDR guarantee in this work is asymptotic and relies on the asymptotic independence of the two transformed p-values. For the same reason, the procedure may be less reliable in small samples, and its finite-sample behavior warrants further investigation. Overall, by leveraging the special structure of high-dimensional mediation testing, the proposed procedure provides an adaptive two-stage approach to large-scale mediation analysis that avoids sample splitting and does not require estimation of the full joint null distribution. We expect this design principle to be useful more broadly in settings where composite null hypotheses give rise to paired p-values.

\section*{Acknowledgements}

During the preparation of this manuscript, ChatGPT (OpenAI) and Claude (Anthropic) were used to assist with language polishing and code debugging. 

\section*{Data Availability}

The data used in Section \ref{sec: data} to support our findings are available as part of the supplementary materials/code archive associated with \cite{liu2022large} through Taylor \& Francis Online. Access is subject to the conditions set by the original source. The code used to reproduce the simulation studies and real-data analysis in this paper is available on GitHub at https://github.com/keyyyyyx/Two-stage-Adaptive-Mediation-Testing.

\bibliographystyle{apalike}
\bibliography{refs}

\clearpage
\appendix
\section*{Supplementary Material}

% Start supplementary numbering
\setcounter{table}{0}
\setcounter{figure}{0}
\setcounter{section}{0}
\renewcommand{\thesection}{S\arabic{section}}
\renewcommand{\thetable}{S\arabic{table}}
\renewcommand{\thefigure}{S\arabic{figure}}

\section*{S1. Proof of Proposition 1}

For notational simplicity, we suppress the subscript $j$ throughout the proof. By the Proposition's assumption, $(p_1,\ p_2)\stackrel{d}{\to}(D_1,\ D_2)$. Since $\min$ and $\max$ are continuous functions, the continuous mapping theorem gives 
$$
    \left(p_{(1)},\ p_{(2)}\right) \stackrel d\to \left(\min(D_1,\ D_2),\ \max(D_1,\ D_2)\right) = \left(D_{(1)},\ D_{(2)}\right)
$$
for the order statistics $p_{(1)}$ and $p_{(2)}$. We first consider the global null, where both $D_1$ and $D_2$ are $Unif(0,1)$ and $D_1$, $D_2$ are independent. The order statistics $\left(D_{(1)},\ D_{(2)}\right)$ have joint density $f(d_1,d_2) = 2$ for $0\leq d_1\leq d_2\leq 1$, and conditioning on $D_{(1)} = d_1$, 
$$
    f_{D_{(2)}|D_{(1)} = d_1}(d_2) = \frac{f(d_1, d_2)}{f_{D_{(1)}}(d_1)} = \frac{2}{2(1-d_1)} = \frac 1{1-d_1},
$$
which implies $D_{(2)}|D_{(1)} = d_1 \sim Unif(d_1, 1)$. Therefore, $1-D_{(2)}|D_{(1)} = d_1 \sim Unif(0, 1-d_1)$, and
$$
    \left.\frac{1-D_{(2)}}{1-D_{(1)}}\right|D_{(1)} = d_1 \sim \frac{Unif(0, 1-d_1)}{1-d_1} = Unif(0, 1).
$$
Since this conditional distribution does not depend on $d_1$, we conclude unconditionally that
$$
    \frac{1-D_{(2)}}{1-D_{(1)}} \sim Unif(0, 1) \ .
$$
In addition, $(1-D_{(1)})^2 \sim Unif(0, 1)$ and is independent of $\frac{1-D_{(2)}}{1-D_{(1)}}$. Therefore,
$$
(p_{1/2}^*,p_{2/2}^*) = \left(1-(1-p_{(1)})^2,1- \frac{1-p_{(2)}}{1-p_{(1)}}\right) \stackrel d\to \left(1-(1-D_{(1)})^2,1- \frac{1-D_{(2)}}{1-D_{(1)}}\right) \sim Unif(0, 1)^{\otimes 2},
$$
where $Unif(0, 1)^{\otimes 2}$ is the standard bivariate uniform distribution.

Under the partial null, $D_{(1)}\equiv 0$, % so $p_{(1)}\stackrel{p}{\to} 0$ and $p^*_{1/2}\stackrel{p}{\to} 0$.  Also, $D_{(2)}\sim Unif(0,1)$ and independent of $D_{(1)}$, so $p^*_{2/2}\stackrel{d}{\to} Unif(0,1)$.  
and $D_{(2)}\sim Unif(0,1)$. Thus $p_{(1)}\stackrel{p}{\to}0$ and
$p_{(2)}\stackrel{d}{\to}D_{(2)}\sim Unif(0,1)$. The continuous mapping theorem gives $p^*_{1/2}=1-(1-p_{(1)})^2\stackrel{p}{\to}0$, and by Slutsky's theorem
\[
  p^*_{2/2}=1-\frac{1-p_{(2)}}{1-p_{(1)}}
  \stackrel{d}{\to}1-\frac{1-D_{(2)}}{1-0}=D_{(2)}\sim Unif(0,1).
\]
Additionally, since $p^*_{1/2}$ converges in probability to the constant $0$, it is asymptotically independent of $p^*_{2/2}$. Under $H_1$, $(D_{(1)},D_{(2)})\equiv (0,0)$. Therefore, $p^*_{1/2}\stackrel{p}{\to} 0$ and $p^*_{2/2}\stackrel{p}{\to} 0$, and $p^*_{1/2},p^*_{2/2}$ are asymptotically independent since both limits are constants.

\section*{S2. Proof of Theorem 2}

Let $\mathbf{P}^{*(n)} = \{(p_{1/2,j}^{*(n)}, p_{2/2,j}^{*(n)})\}_{j=1}^J$ denote the matrix of finite-sample transformed p-values. As $n \to \infty$, by Proposition 2.1,
%under the assumption of asymptotic score orthogonality (Condition C1), 
$\mathbf{P}^{*(n)} \xrightarrow{d} \mathbf{P}^{*(\infty)}$.
Also let $\text{FDP}(\mathbf{P}^*)$ be the mapping from the transformed p-values to the False Discovery Proportion resulting from Algorithm 1. Because Algorithm 1 involves indicator functions, sorting, and data-dependent thresholding, $\text{FDP}(\cdot)$ is a step function with a set of discontinuity points $D_{\text{FDP}}$. These discontinuities occur only on hyperplanes where p-values are exactly tied (altering their rank order) or exactly equal to the data-dependent Stage 1 or Stage 2 thresholds. 

Under the limiting distribution $\mathbf{P}^{*(\infty)}$, the true nulls are continuous and mutually independent; thus, the probability of any ties involving null hypotheses, or of nulls exactly hitting rejection boundaries, is zero. While true non-nulls converge to point masses at $0$ (creating ties among themselves), the $\text{FDP}$ mapping is constant with respect to the internal ranking of these zero values, as they will all trivially fall below the rejection thresholds. Consequently, the limiting distribution assigns zero probability to the effective discontinuities of the mapping: $\mathbb{P}(\mathbf{P}^{*(\infty)} \in D_{\text{FDP}}) = 0$.

By the Extended Continuous Mapping Theorem (\cite{van2000asymptotic}, Theorem 2.3), we have $\text{FDP}(\mathbf{P}^{*(n)}) \xrightarrow{d} \text{FDP}(\mathbf{P}^{*(\infty)})$. Because the FDP is deterministically bounded in $[0,1]$, the Bounded Convergence Theorem ensures that $\lim_{n \to \infty} \mathbb{E}[\text{FDP}^{(n)}] = \mathbb{E}[\text{FDP}^{(\infty)}]$. 

We now establish that $FDR^{(\infty)} =\mathbb{E}[\text{FDP}^{(\infty)}] \le \alpha$. For notational simplicity, we suppress the superscript $(\infty)$ in the following.
Let $t_\gamma$ denote the Stage-1 screening threshold, $S = \{j : p_{1/2,j}^* \leq t_\gamma\}$ denote the set of hypotheses passing Stage 1, $R_1(t_\gamma)=|S|$ be the number of hypotheses passing Stage 1, and $R_2(t_\gamma, t) = |\{j \in S : p_{2/2,j}^* \leq t\}|$ is the number of Stage 2 rejections at threshold $t$. Define the Stage 2 threshold
    $$t_\alpha^\lambda = \sup\left\{t: \widehat {FDR}_\lambda^*(t)\leq\alpha\right\},$$ 
    where 
$$
    \widehat{FDR}_\lambda^*(t)=\begin{cases}\frac{R_1(t_\gamma)\hat\pi_0(\lambda) t}{R_2(t_\gamma,\ t)\vee 1}& \text{if } t \leq \lambda\\ 1 & \text{if } t > \lambda \end{cases},
$$
with 
$$\hat\pi_0(\lambda) = \frac{R_1(t_\gamma)-R_2(t_\gamma,\lambda)+1}{R_1(t_\gamma)(1-\lambda)}, \quad \lambda\in [0, 1)$$
being the Storey's estimator of $\pi_0 = \frac{|J_0\cap S|}{R_1(t_\gamma)}$, the proportion of true nulls among hypotheses passed Stage 1. Here, $J_0$ is the collection of indices of all true null hypotheses. Define filtrations
$$
\mathcal F^{(1)}=\sigma(\mathbb I (p_{1/2,j}^*\leq u),\ j = 1, \dots, J,\ 0\leq u\leq 1) \ ,
$$
$$\mathcal F_t = \sigma(\mathcal F^{(1)},\ \{\mathbb I (p_{2/2,j}^*\leq u), \ j \in S,\ t\leq u\leq 1\}) \ ,$$
$t_\gamma$ and $S$ are adapted to $\mathcal F^{(1)}$ and therefore $\mathcal F_t$.  
Furthermore, let $V(t_\gamma, t) = |\{j \in J_0 \cap S : p_{2/2,j}^* \leq t\}|$ denote the number of false rejections at threshold $t$. For $t\leq s\leq \lambda$, 
$$
% E\left.\left(\frac{V(t_\gamma,t)}{t}\right|\mathcal{F}_t\right)=E\left.\left(\frac{V(t_\gamma,s)}{s}\right|\mathcal{F}_s\right) \ ,
\ E\left.\left(\frac{V(t_\gamma,t)}{t}\right|\mathcal{F}_s\right)=E\left.\left(\frac{V(t_\gamma,s)}{s}\right|\mathcal{F}_s\right)=\frac{V(t_\gamma,s)}{s},
$$
and $t_\alpha^{\lambda}$ is adapted to $\mathcal{F}_t$, so $\frac{V(t_\gamma, t)}{ t}$ is a reverse martingale with respect to $\mathcal F_t$, with starting time $\lambda$ and stopping time $t_\alpha^\lambda$.

Therefore, 
$$FDR = \mathbb E\left[\frac{V(t_\gamma, t_\alpha^\lambda)}{R_2(t_\gamma, t_\alpha^\lambda)}\right] = \mathbb E\left[\frac{V(t_\gamma, t_\alpha^\lambda)}{R_2(t_\gamma, t_\alpha^\lambda)};\ \widehat{FDR}_\lambda^*(\lambda) >\alpha \right]+\mathbb E\left[\frac{V(t_\gamma, t_\alpha^\lambda)}{R_2(t_\gamma, t_\alpha^\lambda)};\ \widehat{FDR}_\lambda^*(\lambda) \leq\alpha \right],$$
where
$$
\begin{aligned}
     &\ \mathbb E\left[\frac{V(t_\gamma, t_\alpha^\lambda)}{R_2(t_\gamma, t_\alpha^\lambda)};\ \widehat{FDR}_\lambda^*(\lambda) >\alpha \right]\\
     \leq&\  \alpha\mathbb E\left[\frac{V(t_\gamma, t_\alpha^\lambda)}{R_1(t_\gamma)\hat\pi_0(\lambda) t_\alpha^\lambda};\ \widehat{FDR}_\lambda^*(\lambda) >\alpha\right] \\
    =&\  \alpha\mathbb E\left\{\frac 1{R_1(t_\gamma)}\left.\mathbb E\left[\frac{V(t_\gamma, t_\alpha^\lambda)}{\hat\pi_0(\lambda) t_\alpha^\lambda}; \ \widehat{FDR}_\lambda^*(\lambda) >\alpha\right|\mathcal F^{(1)}\right]\right\} \\
    =&\  \alpha\mathbb E\left\{\frac 1{R_1(t_\gamma)}\left.\mathbb E\left[\frac{V(t_\gamma, t_\alpha^\lambda)R_1(t_\gamma)(1-\lambda)}{(R_1(t_\gamma)-R_2(t_\gamma,\lambda)+1) t_\alpha^\lambda}; \ \widehat{FDR}_\lambda^*(\lambda) >\alpha\right|\mathcal F^{(1)}\right] \right\} \\
    =&\  \alpha\mathbb E\left\{\left.\mathbb E\left[\frac{V(t_\gamma, t_\alpha^\lambda)(1-\lambda)}{(R_1(t_\gamma)-R_2(t_\gamma,\lambda)+1) t_\alpha^\lambda}; \ \widehat{FDR}_\lambda^*(\lambda) >\alpha\right|\mathcal F^{(1)}\right] \right\}\\
        =&\  \alpha\mathbb E\left\{\left.\mathbb E\left[\frac{1-\lambda}{R_1(t_\gamma)-R_2(t_\gamma,\lambda)+1 }\mathbb E\left.\left(\frac{V(t_\gamma, t_\alpha^\lambda)}{ t_\alpha^\lambda}\right| \mathcal F_{\lambda}\right); \ \widehat{FDR}_\lambda^*(\lambda) >\alpha\right|\mathcal F^{(1)}\right] \right\}\\
            =&\  \alpha\mathbb E\left\{\left.\mathbb E\left[\frac{1-\lambda}{R_1(t_\gamma)-R_2(t_\gamma,\lambda)+1 } \frac{V(t_\gamma, \lambda)}{ \lambda}; \ \widehat{FDR}_\lambda^*(\lambda) >\alpha\right|\mathcal F^{(1)}\right] \right\} \ ,
\end{aligned}
$$
where the last equality follows from the optional stopping theorem.

On the other hand,
$$
\begin{aligned}
    &\ \mathbb E\left[\frac{V(t_\gamma, t_\alpha^\lambda)}{R_2(t_\gamma, t_\alpha^\lambda)};\ \widehat{FDR}_\lambda^*(\lambda) \leq\alpha \right]\\
    =&\ \mathbb E\left[\frac{V(t_\gamma, \lambda)}{R_2(t_\gamma, \lambda)};\ \widehat{FDR}_\lambda^*(\lambda) \leq\alpha \right] \\
    \leq& \ \alpha\mathbb E\left[\frac{V(t_\gamma, \lambda)}{R_1(t_\gamma)\hat\pi_0(\lambda)\lambda};\ \widehat{FDR}_\lambda^*(\lambda) \leq\alpha \right]  \\
    =& \ \alpha\mathbb E\left[\frac{V(t_\gamma, \lambda)}{R_1(t_\gamma)\lambda}\frac{R_1(t_\gamma)(1-\lambda)}{R_1(t_\gamma)-R_2(t_\gamma,\lambda)+1};\ \widehat{FDR}_\lambda^*(\lambda) \leq\alpha \right] \\
    =& \ \alpha\mathbb E\left[\frac{1-\lambda}{R_1(t_\gamma)-R_2(t_\gamma,\lambda)+1}\frac{V(t_\gamma, \lambda)}{\lambda};\ \widehat{FDR}_\lambda^*(\lambda) \leq\alpha \right]  \\
    =& \ \alpha\mathbb E\left\{\left.\mathbb E\left[\frac{1-\lambda}{R_1(t_\gamma)-R_2(t_\gamma,\lambda)+1 } \frac{V(t_\gamma, \lambda)}{ \lambda}; \ \widehat{FDR}_\lambda^*(\lambda) \leq\alpha\right|\mathcal F^{(1)}\right] \right\} .
\end{aligned}
$$
Therefore,
$$
\begin{aligned}
    FDR
    %= \mathbb E\left[\frac{V(t_\gamma, t_\alpha^\lambda)}{R_2(t_\gamma, t_\alpha^\lambda)}\right] = \mathbb E\left[\frac{V(t_\gamma, t_\alpha^\lambda)}{R_2(t_\gamma, t_\alpha^\lambda)};\ \widehat{FDR}_\lambda^*(\lambda) >\alpha \right]+\mathbb E\left[\frac{V(t_\gamma, t_\alpha^\lambda)}{R_2(t_\gamma, t_\alpha^\lambda)};\ \widehat{FDR}_\lambda^*(\lambda) \leq\alpha \right] \\
    &\leq \alpha\mathbb E\left\{\left.\mathbb E\left[\frac{1-\lambda}{R_1(t_\gamma)-R_2(t_\gamma,\lambda)+1 } \frac{V(t_\gamma, \lambda)}{ \lambda}\right|\mathcal F^{(1)}\right] \right\} \\
    &\leq \alpha\mathbb E\left\{\left.\mathbb E\left[\frac{1-\lambda}{R_1(t_\gamma)\pi_0-V(t_\gamma,\lambda)+1 } \frac{V(t_\gamma, \lambda)}{ \lambda}\right|\mathcal F^{(1)}\right] \right\}.
\end{aligned}
$$
The second inequality holds since $R_1(t_\gamma)-R_2(t_\gamma,\lambda)=|\{j: p_{2/2,j}^*>\lambda,\ j\in S\}|$ is at least $R_1(t_\gamma)\pi_0-V(t_\gamma,\lambda)=|\{j: p_{2/2,j}^*>\lambda,\ j\in S\cap J_0\}|$, where $R_1(t_\gamma)\pi_0 = |J_0\cap S|$ is the number of nulls passing Stage 1. Let $N(t_\gamma, \lambda) = R_1(t_\gamma)\pi_0-V(t_\gamma,\lambda)$.  We have
$$
\begin{aligned}
& \mathbb E\left\{\left.\mathbb E\left[\frac{1-\lambda}{R_1(t_\gamma)\pi_0-V(t_\gamma,\lambda)+1 } \frac{V(t_\gamma, \lambda)}{ \lambda}\right|\mathcal F^{(1)}\right] \right\}\\ 
=& \ \mathbb E\left\{\frac{1-\lambda}{\lambda}\left.\mathbb E\left[\frac{V(t_\gamma, \lambda)}{R_1(t_\gamma)\pi_0-V(t_\gamma,\lambda)+1 }\right|\mathcal F^{(1)}\right] \right\} \\
    =&\ \mathbb E\left\{\frac{1-\lambda}{\lambda}\left.\mathbb E\left[\frac{R_1(t_\gamma)\pi_0-N(t_\gamma, \lambda)}{N(t_\gamma, \lambda)+1 }\right|\mathcal F^{(1)}\right] \right\}\\
    =& \ \mathbb E\left\{\frac{1-\lambda}{\lambda}\left.\mathbb E\left[R_1(t_\gamma)\pi_0\frac{1}{N(t_\gamma, \lambda)+1 }-\frac{N(t_\gamma, \lambda)}{N(t_\gamma, \lambda)+1}\right|\mathcal F^{(1)}\right] \right\} \\
    =& \ \mathbb E\left\{\frac{1-\lambda}{\lambda}\left.\mathbb E\left[(R_1(t_\gamma)\pi_0+1)\frac{1}{N(t_\gamma, \lambda)+1 }-1\right|\mathcal F^{(1)}\right] \right\}\\
    =& \ \mathbb E\left\{\frac{1-\lambda}{\lambda}\left[\left.(R_1(t_\gamma)\pi_0+1)\mathbb E\left[\frac{1}{N(t_\gamma, \lambda)+1 }\right|\mathcal F^{(1)}\right]-1\right] \right\}
    \\
    =& \ \mathbb E\left\{\frac{1-\lambda}{\lambda}\left[(R_1(t_\gamma)\pi_0+1)\frac{1-\lambda^{R_1(t_\gamma)\pi_0+1}}{(R_1(t_\gamma)\pi_0+1)(1-\lambda)}-1\right] \right\}\\
    =& \ \mathbb E\left\{\frac{1-\lambda}{\lambda}\left[\frac{1-\lambda^{R_1(t_\gamma)\pi_0+1}}{1-\lambda}-1\right] \right\} \\
    =&  \ \mathbb E\left\{1-\lambda^{R_1(t_\gamma)\pi_0} \right\} \leq 1 \ .
\end{aligned}
$$
The sixth equality holds since under the independence between $(p_{1/2,j}^*,\ p_{2/2,j}^*)$ and mutual independence across null $p_{2/2,j}^*$'s, $N(t_\gamma,\lambda)|\mathcal{F}^{(1)}$ is $Binom(R_1(t_\gamma)\pi_0,1-\lambda)$. Therefore, $\mathbb{E}[\text{FDP}^{(\infty)}] \le \alpha$.

\section*{S3. Proof of Remark 3}

We establish $\mathbb{E}[\text{FDP}^{(\infty)}] \le \alpha$, and asymptotic control follows the same argument as in S2.  Let $t_\gamma$ and $t_\alpha = \sup\left\{t:\frac{R_1(t_\gamma)\hat\pi_0 t}{R_2(t)\vee 1}\leq\alpha\right\}$ denote the thresholds for Stage 1 and Stage 2, respectively, where $\hat\pi_0$ is an arbitrary estimate of $\pi_0 = \frac{|J_0\cap S|}{R_1(t_\gamma)}$, $J_0$ is the collection of indices of all true null hypotheses, and $S=\{j: p_{1/2, j}^*\leq t_\gamma\}$ is the set of indices of hypotheses passed Stage 1. Note that $|S| = R_1(t_\gamma)$. Then 
$$
\begin{aligned}
    FDR &= \mathbb E\left[\frac{V(t_\gamma, t_\alpha)}{R_2(t_\gamma, t_\alpha)}\right]\leq  \alpha\mathbb E\left[\frac{V(t_\gamma, t_\alpha)}{R_1(t_\gamma)\hat\pi_0 t_\alpha}\right] \\
    &= \alpha\mathbb E\left\{\frac 1{R_1(t_\gamma)}\left.\mathbb E\left[\frac{V(t_\gamma, t_\alpha)}{\hat\pi_0 t_\alpha}\right|\mathcal F^{(1)}\right] \right\}.
\end{aligned}
$$
Let $\mathbf p$ denote the vector containing $p_{2/2, j}^*$ for all $j\in S$.  By the assumption that the estimator $\hat\pi_0 = \hat\pi_0\left(\mathbf p\right):[0, 1]^{|S|}\mapsto [0, 1]$ is coordinate-wise non-decreasing in the Stage 2 $p$-values, $1/\hat\pi_0(\mathbf p)$ is coordinate-wise non-increasing. For each $j\in S$, let $\mathbf p_{-j}$ denote $\mathbf p$ with $p_{2/2, j}^*$ removed, and let $\mathbf p_{0, j}$ denote $\mathbf p$ with $p_{2/2, j}^*$ being replaced by $0$. Note that $\hat\pi_0(\mathbf p_{0, j})\leq\hat\pi_0(\mathbf p)$ for all $j\in S$. Thus we have
$$
    \begin{aligned}
    &\, \left.\mathbb E\left[\frac{V(t_\gamma, t_\alpha)}{\hat\pi_0 t_\alpha}\right|\mathcal F^{(1)}\right] =\sum_{j\in J_0} \left.\mathbb E\left[\frac{\mathbb I\left(p_{1/2,j}^*\leq t_\gamma,\ p^*_{2/2,j}\leq t_\alpha\right)}{\hat\pi_0(\mathbf p) t_\alpha}\right|\mathcal F^{(1)}\right] \\
    & =\sum_{j\in J_0\cap S} \left.\mathbb E\left[\frac{\mathbb I\left(p^*_{2/2,j}\leq t_\alpha\right)}{\hat\pi_0(\mathbf p) t_\alpha}\right|\mathcal F^{(1)}\right] \leq \sum_{j\in J_0\cap S} \left.\mathbb E\left[\frac{\mathbb I\left(p^*_{2/2,j}\leq t_\alpha\right)}{\hat\pi_0(\mathbf p_{0,j}) t_\alpha}\right|\mathcal F^{(1)}\right] \\
    &= \sum_{j\in J_0\cap S} \left.\mathbb E\left\{\left.\mathbb E\left[\frac{\mathbb I\left(p^*_{2/2,j}\leq t_\alpha\right)}{\hat\pi_0(\mathbf p_{0,j}) t_\alpha}\right| \textcolor{black}{\sigma\left(\mathbf p_{-j}, \mathcal F^{(1)}\right)}\right]\right|\mathcal F^{(1)}\right\} \\
    &= \sum_{j\in J_0\cap S} \left.\mathbb E\left\{\frac 1{\hat\pi_0(\mathbf p_{0,j})}\left.\mathbb E\left[\frac{\mathbb I\left(p^*_{2/2,j}\leq t_\alpha\right)}{ t_\alpha}\right|\textcolor{black}{\sigma\left(\mathbf p_{-j}, \mathcal F^{(1)}\right)}\right]\right|\mathcal F^{(1)}\right\} \ . 
\end{aligned}
$$
In addition, for each $j\in J_0\cap S$, $\frac{\mathbb I(p_{2/2, j}^*\leq t)}{t}$ is a reverse martingale with respect to $\mathcal F_{t}$, and $t_{\alpha}$ is a stopping time. By the optional stopping theorem, 
$$
    \left.\mathbb E\left[\frac{\mathbb I\left(p^*_{2/2,j}\leq t_\alpha\right)}{ t_\alpha}\right|\textcolor{black}{\sigma\left(\mathbf p_{-j}, \mathcal F^{(1)}\right)}\right] = \left.\mathbb E\left[\frac{\mathbb I\left(\ p^*_{2/2,j}\leq 1\right)}{ 1}\right|\textcolor{black}{\sigma\left(\mathbf p_{-j}, \mathcal F^{(1)}\right)}\right] = 1
$$
for each $j\in J_0\cap S$. Therefore, we have
$$
    \begin{aligned}
    &\, \left.\mathbb E\left[\frac{V(t_\gamma, t_\alpha)}{\hat\pi_0 t_\alpha}\right|\mathcal F^{(1)}\right] \leq \sum_{j\in J_0\cap S} \left.\mathbb E\left\{\frac 1{\hat\pi_0(\mathbf p_{0,j})}\right|\mathcal F^{(1)}\right\},
\end{aligned}
$$
and hence
$$
    \begin{aligned}
        FDR \leq \alpha\mathbb E\left\{\frac 1{R_1(t_\gamma)}\sum_{j\in J_0\cap S} \left.\mathbb E\left[\frac 1{\hat\pi_0(\mathbf p_{0,j})}\right|\mathcal F^{(1)}\right] \right\}.
    \end{aligned}
$$
If $\hat\pi_0$ is conservative, in particular, $\mathbb E\left.\left[\frac 1{\hat\pi_0(\mathbf p_{0,j})}\right|\mathcal F^{(1)}\right] \leq \pi_0^{-1}$, we then have
$$
FDR \leq \alpha\mathbb E\left\{\frac 1{R_1(t_\lambda)}\frac{|J_0\cap S|}{\pi_0}\right\} = \alpha\mathbb E\left\{\frac 1{R_1(t_\lambda)}\frac{\pi_0 R_1(t_\lambda)}{\pi_0}\right\} = \alpha.
$$
%(Proof referred to \citealp{STS_2004} Theorem 2, \citealp{blanchard2009adaptive} Theorem 11, and \citealp{li2025note} Theorem 1).

\section*{S4. Sensitivity Analysis of Bounding Sequence $c_m$}

The simulation-based calibration of the bounding sequence $c_J^{\mathrm{sim}}$ in the AdSMR procedure accounts for dependence among hypotheses but requires drawing a large number of null replicates, making it computationally intensive when $J$ is large. In contrast, the closed-form Gumbel approximation $c_J^{\mathrm{ind}}$ derived in \cite{meinshausen2006estimating} relies on an independence assumption but avoids the simulation step entirely, and is therefore computationally much less demanding. 

\begin{figure}
  \centering
  \includegraphics[width=\textwidth, alt = {Three scatter plots side by side, one for each dependence block size (b equal to 10, 50, and 100). Each plots the independence-approximated signal-proportion estimator (vertical axis) against the simulation-based estimator (horizontal axis), both ranging from about 0.02 to 0.10, with the number of hypotheses fixed at 20,000. Point color denotes the within-block correlation (0.3, 0.5, and 0.7). Points cluster around the three true signal proportions of 0.02, 0.05, and 0.10, marked by black crosses, and lie slightly above the dashed diagonal equality line, indicating the independence-based estimator is slightly larger. The gap above the line widens as block size and correlation increase but stays small in every panel.}]{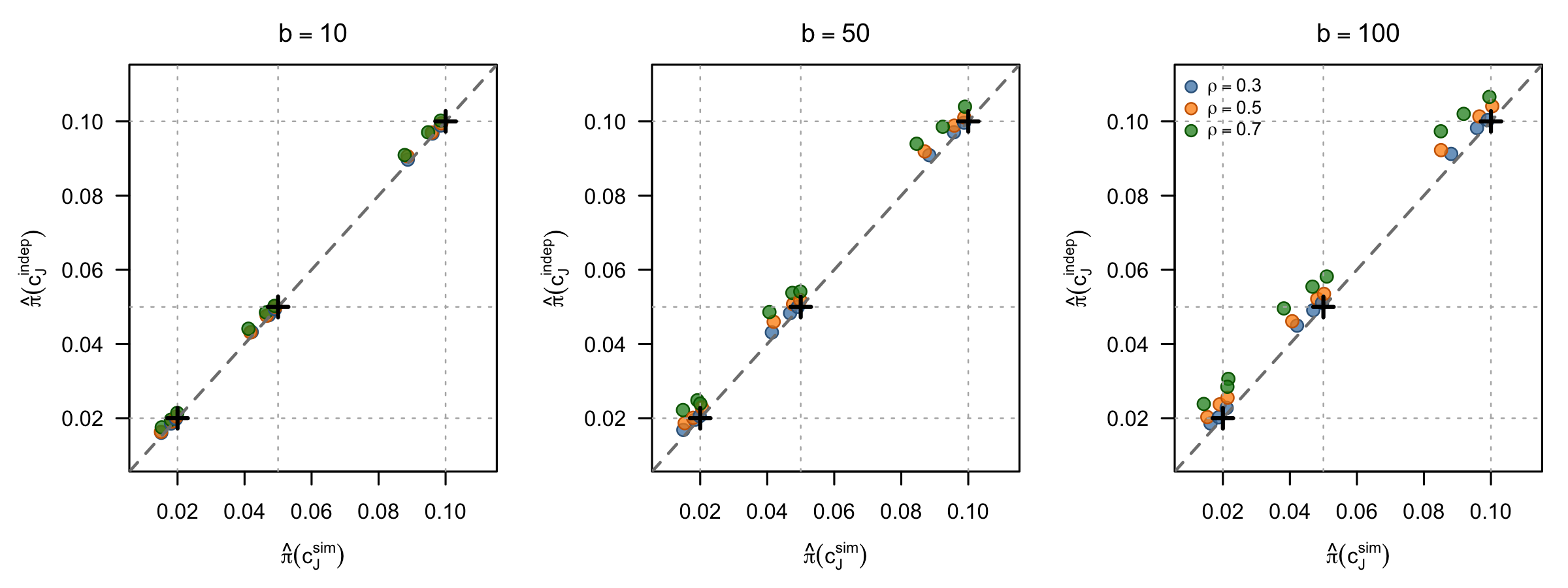}
  \caption{Scatter plot of $\hat{\pi}(c_J^{\mathrm{ind}})$ versus
    $\hat{\pi}(c_J^{\mathrm{sim}})$ across simulation configurations with number of hypotheses fixed at $J=20,000$, faceted by dependence block size $b$. Colors indicate the within-block correlation $\rho$. The dashed diagonal line marks the $\hat{\pi}(c_J^{\mathrm{ind}})=\hat{\pi}(c_J^{\mathrm{sim}})$ line; black crosses mark the true signal proportions $\pi \in \{0.02, 0.05, 0.10\}$. Points consistently lie above the diagonal, confirming that the independence-based approximation yields a slightly larger $\hat{\pi}_{\mathrm{MR}}$. Although the discrepancy grows with block size and correlation strength, it remains small across all configurations.}
  \label{fig:pihat-comparison}
\end{figure}

\begin{figure}
  \centering
  \includegraphics[width=\textwidth, alt = {Three line charts side by side, one for each block size (b equal to 10, 50, and 100), all at a signal proportion of 0.05. Each plots the mean absolute difference between the simulation-based and independence-approximated estimators (vertical axis, from 0 to about 0.015) against the number of hypotheses (horizontal axis, from 5,000 to 20,000). Within each panel, three lines correspond to within-block correlations of 0.3, 0.5, and 0.7. In every panel, the difference decreases as the number of hypotheses grows and is larger for higher correlation; the overall magnitude increases with block size, reaching about 0.014 for correlation 0.7 at block size 100, while remaining small throughout.}]{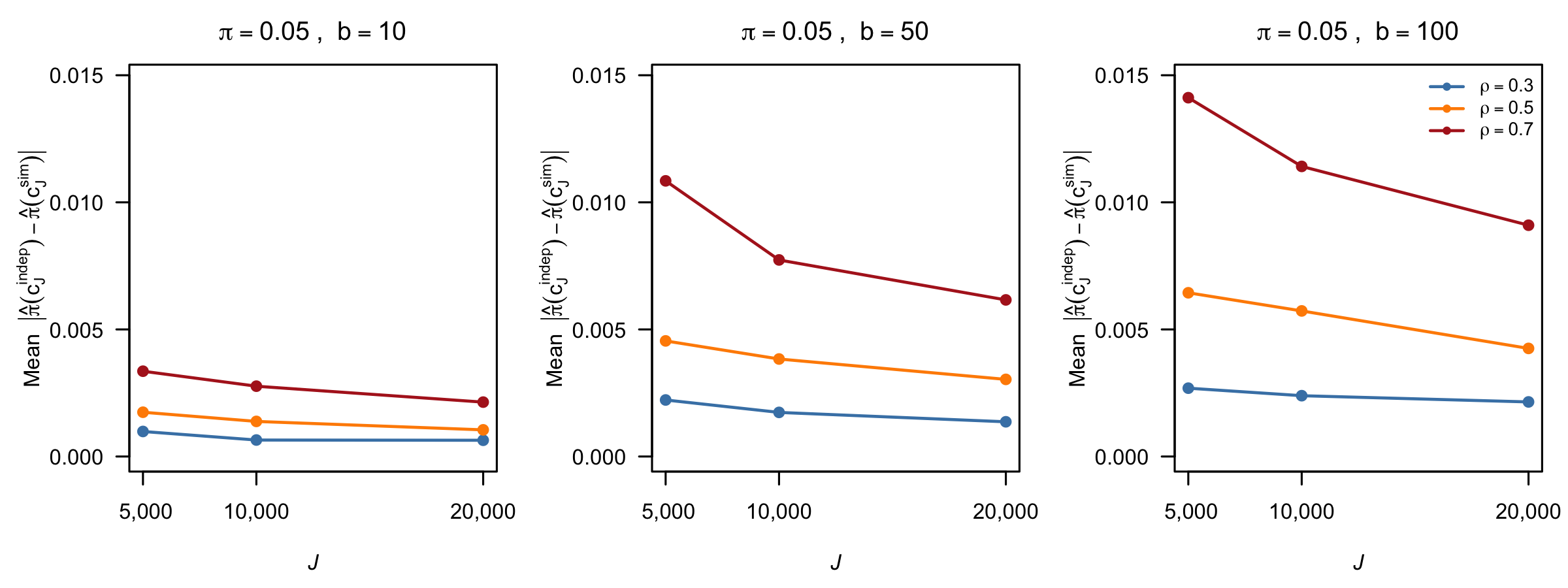}
  \caption{Mean absolute difference between the simulation-based estimator $\hat{\pi}(c_J^{\mathrm{sim}})$ and the independence-approximated estimator $\hat{\pi}(c_J^{\mathrm{ind}})$ as a function of the number of hypotheses $J$, with $\pi = 0.05$ and panels faceted by block size $b$. Within each panel, lines correspond to different within-block correlations $\rho$.}
  \label{fig:pihat-convergence}
\end{figure}

To assess the practical impact of replacing $c_J^{\mathrm{sim}}$ with $c_J^{\mathrm{ind}}$, we conducted a sensitivity analysis across configurations varying the number of hypotheses $J\in\{5000, 10000, 20000\}$, block size $b \in\{10, 50, 100\}$, within-block correlation $\rho \in \{0.3, 0.5, 0.7\}$, and true signal proportion $\pi\in\{0.02, 0.05, 0.10\}$. Figure \ref{fig:pihat-comparison} shows that $\hat{\pi}(c_J^{\mathrm{ind}})$ tracks $\hat{\pi}(c_J^{\mathrm{sim}})$ closely across all settings for $J = 20,000$: all points lie just above the diagonal, confirming the expected mild upward shift under the independence approximation, with larger discrepancies observed for larger block sizes and stronger correlations. Figure \ref{fig:pihat-convergence} further quantifies this gap, showing that the mean absolute difference $|\hat{\pi}(c_J^{\mathrm{ind}})-\hat{\pi}(c_J^{\mathrm{sim}})|$ decreases consistently as $J$ increases, and stays below $0.015$ even in the most extreme configuration considered ($b=100$, $\rho = 0.7$). Together, these results suggest that the independence approximation provides a reliable and computationally efficient substitute for the simulation-based bounding sequence, particularly when the number of hypotheses is large.

\section*{S5. Alternative Stage 1 Threshold Selection Rule}

\begin{figure}
  \centering
  \includegraphics[width=0.8\textwidth, alt = {A two-by-two grid of line charts showing the Stage 1 F1-score (vertical axis) against the screening threshold t-gamma (horizontal axis, from 0 to 1), for signal proportions of 0.01 (upper left), 0.04 (upper right), 0.1 (lower left), and 0.2 (lower right). In each panel a single curve rises to a peak at a small threshold and then declines as the threshold increases, with the peak F1-score higher for larger signal proportions (from about 0.73 at 0.01 to about 0.9 at 0.2). A vertical dashed line marks the threshold selected by Algorithm 3, which sits at or near the F1-score peak in every panel.}]{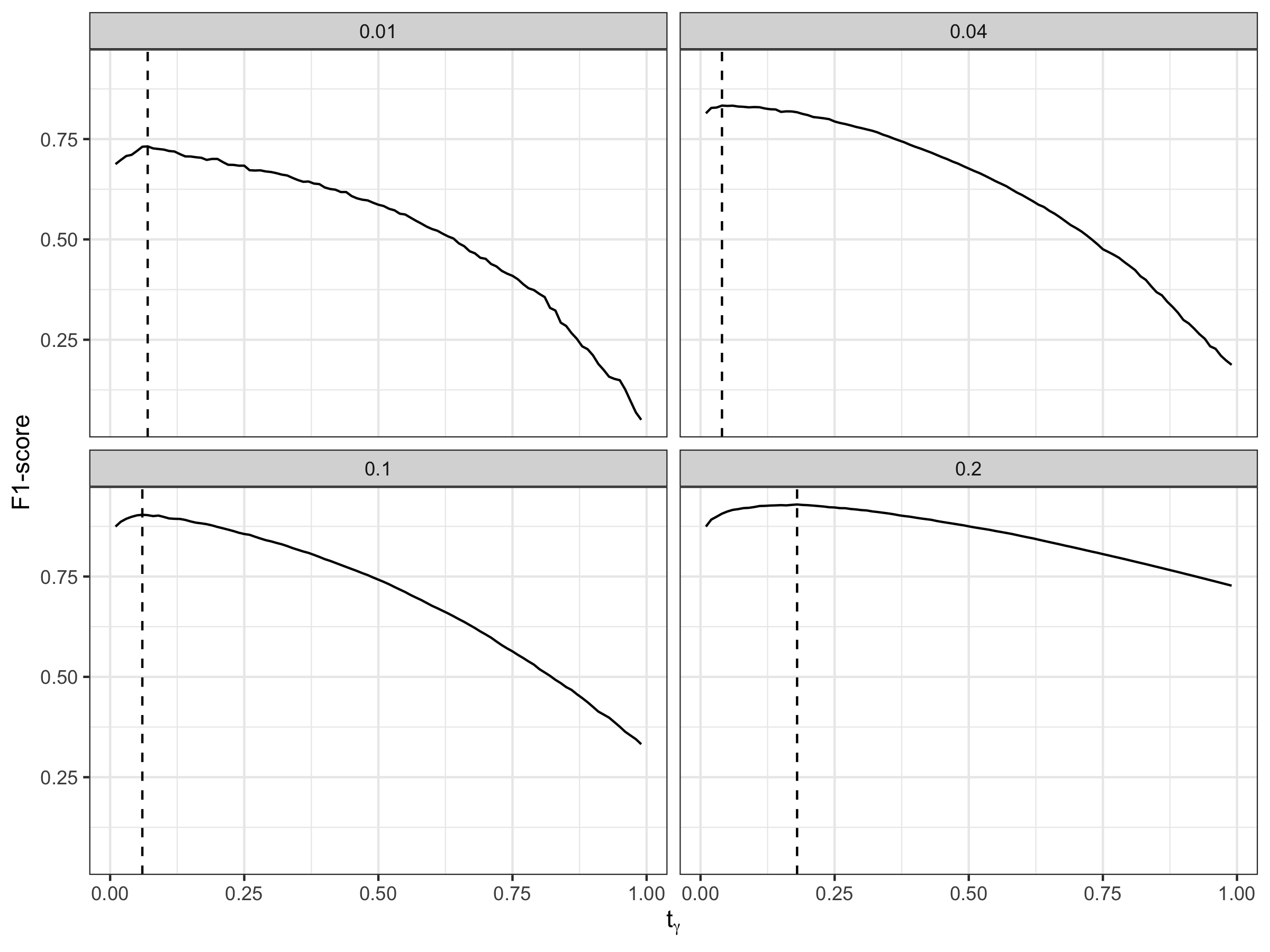}
  \caption{Stage 1 F1-score at different values of $t_\gamma$, under signal proportions 0.01 (upper left), 0.04 (upper right), 0.1 (lower left), and 0.2 (lower right). In each panel, the vertical dashed line marks the Stage 1 threshold $t_\gamma$ selected by Algorithm \ref{alg: F1-score}.}
  \label{fig: F1_vs_gamma}
\end{figure}

% \textcolor{blue}{[include in main or supplement?]} 
A possible alternative is to select $t_\gamma$ that maximizes Stage 1 F1-score
$$F_1 = \frac{2TP}{2TP+FP+FN}=\frac{2TP}{R_1+P_1},$$
where $P_1 = TP+FN$ is the number of non-null hypotheses. Here, only hypotheses with joint nulls are treated as negatives; all other hypotheses with partial nulls or joint alternatives are treated as positive. Maximizing F1-score encourages a large TP while penalizing both false positives and false negatives, thereby providing a data-driven trade-off between (i) retaining as many potentially non-null candidates as possible and (ii) limiting the number of joint-null hypotheses that are carried forward to Stage 2. % \textcolor{blue}{[add a figure demonstrating F1-score as a function of $t_\gamma$]}. 
Figure \ref{fig: F1_vs_gamma} demonstrates how Stage 1 F1-score varies with $t_\gamma$. The Stage 1 F1-score is maximized at relatively small values of $t_\gamma$ in the sparse and moderately sparse settings ($\pi_{11} = 0.01,\ 0.04,\ $ and $0.1$), whereas the optimal threshold shifts to a larger value when signals become denser ($\pi_{11} = 0.2$). Across all panels, the F1-score curve initially increases when $t_\gamma$ is very small, where screening is overly conservative and removes true signals. When $t_\gamma$ gets large, F1-score decreases quickly as too many global nulls are allowed to pass into Stage 2 and offsetting the benefit of retaining a few additional nonnulls. The dashed vertical line marks the values of $t_\gamma$ that maximize Stage 1 F1-score, which balances the two competing goals of Stage 1: retaining as many alternatives as possible while filtering out as many global nulls as possible. The detailed selection rule is outlined in Algorithm \ref{alg: F1-score}.  In step 3, because the tail false discovery rate estimates the probability that a rejected hypothesis is a false positive, multiplying the number of rejections by 1 minus the tail FDR yields the estimated number of true positives.

\begin{algorithm}
\setcounter{algorithm}{2}
\caption{Selecting $t_\gamma$ by maximizing F1-score} 
\label{alg: F1-score}
Let $\Gamma = \{t_{\gamma_1}, \dots,t_{\gamma_k}\}$ denote the set of $t_\gamma$ candidates, e.g. $\Gamma = \{0.05, 0.1, \dots, 0.95\}$.
\begin{enumerate}
    \item Estimate the total number of nonnulls in Stage 1 as $\hat P_1 = J\cdot (1-\hat\pi_{00})$, where $\pi_{00}$ is the proportion of joint nulls, and $\hat\pi_{00}$ can be any proportion estimator, e.g., the Meinshausen–Rice estimator or Storey's estimators. 
    \item For each $t_{\gamma_i}\in \Gamma$, let $I(t_{\gamma_i})=\{j: p_{1/2,\ j}^*\leq t_{\gamma_i}\}$ be the index set of hypotheses passed Stage 1 if using $t_{\gamma_i}$ as rejection threshold, and $R_1(t_{\gamma_i}) = \left|I(t_{\gamma_i})\right|$ be the corresponding number of hypotheses passed Stage 1. 
    \item For each $t_{\gamma_i}\in \Gamma$, estimate Stage 1 TP as $\hat {TP}(t_{\gamma_i}) = R_1(t_{\gamma_i})\left(1-\hat{\text{Fdr}}\left(\max_{j\in I(t_{\gamma_i})}\{p_{1/2,\ j}^*\}\right)\right)$, where $\text{Fdr}(p)=\mathbb P(\text{null}|P\leq p)$ is the tail FDR, and $\max_{j\in I(t_{\gamma_i})}\{p_{1/2,\ j}^*\}$ is the greatest $p$-value corresponding to the set of hypotheses passed Stage 1. Then $\hat{\text{Fdr}}\left(\max_{j\in I(t_{\gamma_i})}\{p_{1/2,\ j}^*\}\right) = \hat{\mathbb P}(\text{null}|\text{rejected})$ estimates the probability of making a false discovery (\citealp{strimmer2008unified}). 
    \item Set $t_\gamma = \arg\max_{t_{\gamma_i}\in\Gamma}\left\{\hat F_1(t_{\gamma_i})\right\}= \arg\max_{t_{\gamma_i}\in\Gamma}\left\{\frac{2\hat{TP} (t_{\gamma_i})}{R_1(t_{\gamma_i})+\hat P_1}\right\}.$
\end{enumerate}
\end{algorithm}

\end{document}